# Optimal transmission expansion modestly reduces decarbonization costs of U.S. electricity *

Rangrang Zheng,[†] Greg Schivley,[‡] Matthias Fripp,[§] Michael J. Roberts[¶]

**Abstract**

Major government studies and policy reports project that substantial expansion of interregional transmission will be needed to integrate clean energy and ensure reliability in decarbonized power systems (DOE, 2023; U.S. Department of Energy, 2024). Using the open-source Switch capacity expansion model with detailed representation of existing U.S. generation and transmission infrastructure, solar, wind, and storage resources, and hourly operations, we evaluate the role of interregional transmission across least-cost, carbon-priced, and zero-emissions scenarios for 2050. An optimal nationwide plan would more than triple interregional transmission capacity, yet this reduces the cost of a zero-emissions system by only 7% relative to relying on existing interregional transmission, as storage, solar and wind siting, and nuclear generation serve as close substitutes. Regional cost and rent effects vary, with transmission generally favoring wind and hydrogen resources over solar and batteries. Sensitivity analysis shows diminishing returns: one-fifth of the benefits of full expansion can be achieved with one-twelfth of the added capacity, while cost reductions for batteries and hydrogen provide comparable or greater system savings than interregional transmission. Upgrading existing interregional corridors with advanced conductors—roughly doubling capacity per link at half the cost of new builds—reduces system costs by only 1.6%, suggesting that reconductoring's benefits are modest and that realizing their full potential likely requires pairing with new connections on key corridors or complementary reductions in battery costs. These results suggest that while substantial transmission expansion is economically justified, a diverse set of flexibility resources can substitute for large-scale grid build-out, and the relative value of transmission is highly contingent on technological and cost developments.

---

*First version: February 21, 2024. This version: May 28, 2026. The authors would like to acknowledge all contributors to Switch 2.0, the open-source capacity expansion model we use in this work, as well as Catalyst Cooperative's Public Utility Data Liberation (PUDL) Project (`https://catalyst.coop/pudl/`), which is a key input into PowerGenome. Partial funding for this project was provided by the Sloan Foundation via the Environmental Defense Fund's Model Intercomparison Project. Comments welcome.

[†]University of Hawai'i, Email: zhengr@hawaii.edu

[‡]Princeton University, Email: greg.schivley@princeton.edu

[§]Energy Innovation: Policy & Technology LLC, Email: matthias@energyinnovation.org

[¶]University of Hawai'i, Email: mjrobert@hawaii.edu

# 1 Background

Although solar and wind power are increasingly affordable and are likely to be central to global decarbonization efforts, their variability presents a considerable challenge to existing systems, which are primarily designed around easily controllable thermal power plants. Interregional transmission can mitigate these challenges by balancing supply and demand across regions and facilitating the transfer of electricity from areas rich in solar and wind resources to those with fewer renewable options. For instance, the desert Southwest receives more solar radiation than other regions, while the Great Plains have abundant wind resources. These aspects of solar and wind contrast sharply with thermal power plants, which perform nearly the same wherever they are placed. For example, the NREL Interconnections Seam Study finds that expanding transmission across U.S. interconnections can significantly reduce system costs, improve reliability, and lower renewable curtailment. However, such studies primarily focus on engineering and system-level outcomes and provide limited insight into the economic value of transmission relative to alternative flexibility options. Because much of the existing interregional transmission infrastructure was not built with a high-renewable future in mind, it remains uncertain whether current transmission lines are well-suited for the shifting interregional and intertemporal dynamics of an increasingly renewable-powered grid. At the same time, extreme weather events and reliability impacts (e.g., heat waves and wildfires in California and winter storms in Texas) have increased discussion of the need for strong transmission connections between U.S. grids.(Botterud et al., 2024; DOE, 2023; U.S. Department of Energy, 2024).

Despite its potential benefits, expanding interregional transmission capacity is fraught with economic, logistical, environmental, and political obstacles. The construction of new transmission lines is historically contentious due to disputes over siting, financing mechanisms, and cost allocation among stakeholders. While increased interregional trade generally improves overall efficiency, some stakeholders—including existing transmission line owners and locally advantaged generators—may face economic losses due to heightened competition (Hausman, 2024). Additionally, transmission projects often traverse regions that do not directly benefit from them, and these communities may understandably resist their siting, particularly when such projects pose environmental and biodiversity risks (Biasotto and Kindel, 2018; Marshall and Baxter, 2002; Söderman, 2006; Hyde, Bohlman and Valle, 2018). Furthermore, ERCOT, the isolated Texas grid, is particularly resistant to interconnections due to its unique market structure and longstanding aversion to Federal Energy Regulatory Commission (FERC) oversight. These challenges raise doubts about whether interregional transmission expansion will progress in line with conventional assessments of its benefits and costs.

Previous research has highlighted the key role of transmission expansion in decarbonization (Pacala et al., 2021; Moch and Lee, 2022; Joskow, 2020; Davis, Hausman and Rose, 2023) in the US and Europe(Golombek et al., 2022), as well as explored various expansion strategies (Rosellón, 2003; Olmos, Rivier and Pérez-Arriaga, 2018). However, no studies have provided a comprehensive assessment of the economic value of interregional transmission expansion in achieving a zero-emission electricity system conditional on the existing U.S. system. Such an assessment requires a thorough consideration of substitutes and complements of transmission, and how different stakeholders and customers could be affected by transmission expansion and its alternatives. To the best of our knowledge, the study by Brown and Botterud (2021) is the only work to thoroughly analyze transmission in a 100% renewable future of the US electricity system using a capacity expansion model. Their results suggest that upgrading interregional transmission infrastructure could reduce the cost of decarbonized power by approximately 30%, lowering the average electricity price from over $100 per MWh to about $70 per MWh. However, a key limitation of their analysis is its focus on an idealized "greenfield" transmission system, with little attention to existing infrastructure. Moreover, while their model features high temporal resolution, it does not fully account for seasonal balancing solutions, such as green hydrogen or carbon capture and storage (CCS) paired with natural gas. The potential deployment of these technologies—despite their costs—could significantly alter decarbonization pathways and reduce the economic value of additional transmission capacity. In the Northeastern North American context, Rodríguez-Sarasty, Debia and Pineau (2021) demonstrate that deep decarbonization can be achieved more cost-effectively when electricity market integration is strengthened and hydropower resources are leveraged as a source of long-duration flexibility. Following this line of inquiry, Ba, Caron and Pineau (2024) analyze whether storage or transmission provides greater value for decarbonization. They show that optimal transmission expansion reduces total system costs by 9%, primarily by lowering required generation capacity by 6% since abundant hydropower acts as a natural storage buffer. These findings reinforce the widespread view that transmission investment is beneficial, but more research is needed to flesh out the range of substitution possibilities with scenarios that explore a wide range of alternative solutions and their costs.

Our study advances the literature by employing a capacity-expansion framework similar in structure to those cited above—jointly optimizing generation, storage, and interregional transmission subject to hourly chronological operational constraints—while extending the analysis to a more comprehensive treatment of the substitutes and complements of transmission. Specifically, we assess interregional transmission value across least-cost, zero-emissions, and carbon-priced scenarios while building from the existing infrastructure base. Our model captures how various technologies—including renewables, batteries, hydrogen, nuclear, and CCS—can complement or substitute for interregional transmission expansion. It further accounts for rents earned by high-capacity or well-timed solar and wind resources, as well as rents earned by pre-existing resources. Using high temporal and spatial resolution, we present a framework that reflects weather-driven variability in electricity supply and demand, providing a comprehensive assessment of the role of interregional transmission and its available substitutes in a decarbonized energy system.

Our main finding is that an optimal zero-emissions plan for U.S. power systems would more than triple current interregional transmission, confirming the conventional wisdom that far more capacity is needed. Yet such a system would cost only 7% less than a well-designed zero-emissions system with *no* new interregional transmission, because a diverse set of substitutes—from storage to adjustments in the generation mix—can contain excess costs, provided these other resources are not themselves constrained. We test the robustness of these findings using a proxy to account for reserves or extreme weather by boosting demand 25% above projections on the three consecutive days in each region that are the most costly to balance, which raises transmission value to 7.7%. We perform many other sensitivity analyses to examine how transmission varies under different decarbonization scenarios and different assumptions about costs and availability of key resources, including batteries, transmission, and demand response.

Our cost reduction estimates differ from others in the literature. For example, the 9% cost reduction reported by Ba, Caron and Pineau (2024) arises in a hydro-abundant context, where low-cost power and long-duration reservoir storage complement transmission. Unlike batteries, however, hydrological generation and storage are geographically constrained. In contrast, the relatively hydro-scarce U.S. system, clean energy solutions must pair solar and wind with a delicate mix of batteries, hydrogen, nuclear, and CCS. Optimal solutions can be sensitive to the mix of resources available in different regions. These differences underscore that the value of transmission depends both on the cost metric employed and on regional resource endowments. Relatedly, transmission expansion raises rents for some resources and regions while lowering them for others. In general, transmission expansion tends to favor producers in regions with high-quality wind or solar resources and consumers in regions that lack them.

Our work contributes to the literature in three ways. First, we characterize and evaluate substitutes for optimal transmission expansion in high-renewable systems. Specifically, we assess transmission value in least-cost, zero-emissions, and carbon-priced scenarios while building from the existing infrastructure base, including transmission. This analysis accounts for a broad range of technologies—renewables, batteries, hydrogen, nuclear, and carbon capture and storage (CCS)—that can substitute for transmission, using a model with high temporal and spatial resolution of weather and associated supply and demand. Second, we show how both transmission value and the overall costs of decarbonization vary as the costs of critical flexibility resources change, including those of transmission, batteries, hydrogen, and demand response. Third, we incorporate rents to scarce and pre-existing resources, which are obscured in conventional levelized-cost analyses, and evaluate the distribution of gains and losses to owners of existing transmission and other generation resources under optimal expansion.

Recent work has continued to build understanding of transmission's role. Botterud et al. (2025) evaluate the proposed BIG WIRES Act using the GenX capacity expansion model, finding that mandated minimum interregional transfer capability requirements yield meaningful cost savings and emissions reductions under deep decarbonization, with additional reliability benefits during extreme weather events analogous to Winter Storm Uri. Their focus is on a specific policy instrument and on reliability; our work complements this by comprehensively mapping the substitution possibilities that determine transmission's relative value across a wide range of technology cost scenarios, and by incorporating the distribution of economic rents that shapes the political economy of transmission expansion. Schivley et al. (2025) present a rigorous multi-model intercomparison–including both Switch (our model) and GenX–confirming that these models produce consistent results when given harmonized inputs and data, which facilitates comparison across studies.

## 2 Current and idealized future systems

The value of expanding transmission depends on what already exists, including generation and transmission infrastructure. It also depends on the resource availability to build and connect new generation and on expected future demand. The existing grid evolved from primarily local power provision to increasingly interconnected regions. At first, transmission was used mainly to transport power from remote generating plants to urban customers (Decker, 2021). Over time, it became clear that efficiency would be enhanced by connecting population centers to exploit economies of scale with large “base load" power plants that could operate continuously at full capacity (e.g., coal and nuclear), paired with smaller local ramping and peaking power plants that could accommodate time-varying demand (Masters, 2013). A larger, more connected grid also has more inertia and stability in the face of demand shocks, severe weather events, plant outages, and, increasingly, renewable energy intermittency. As a result of these economic forces, regions are more interconnected today than a century ago. Still, the U.S. grid remains highly fragmented, with three nearly disjoint sections, the Eastern, Western, and ERCOT interconnections (panel **b** of Figure 1), and varying degrees of connectivity between regions within each of these interconnects.

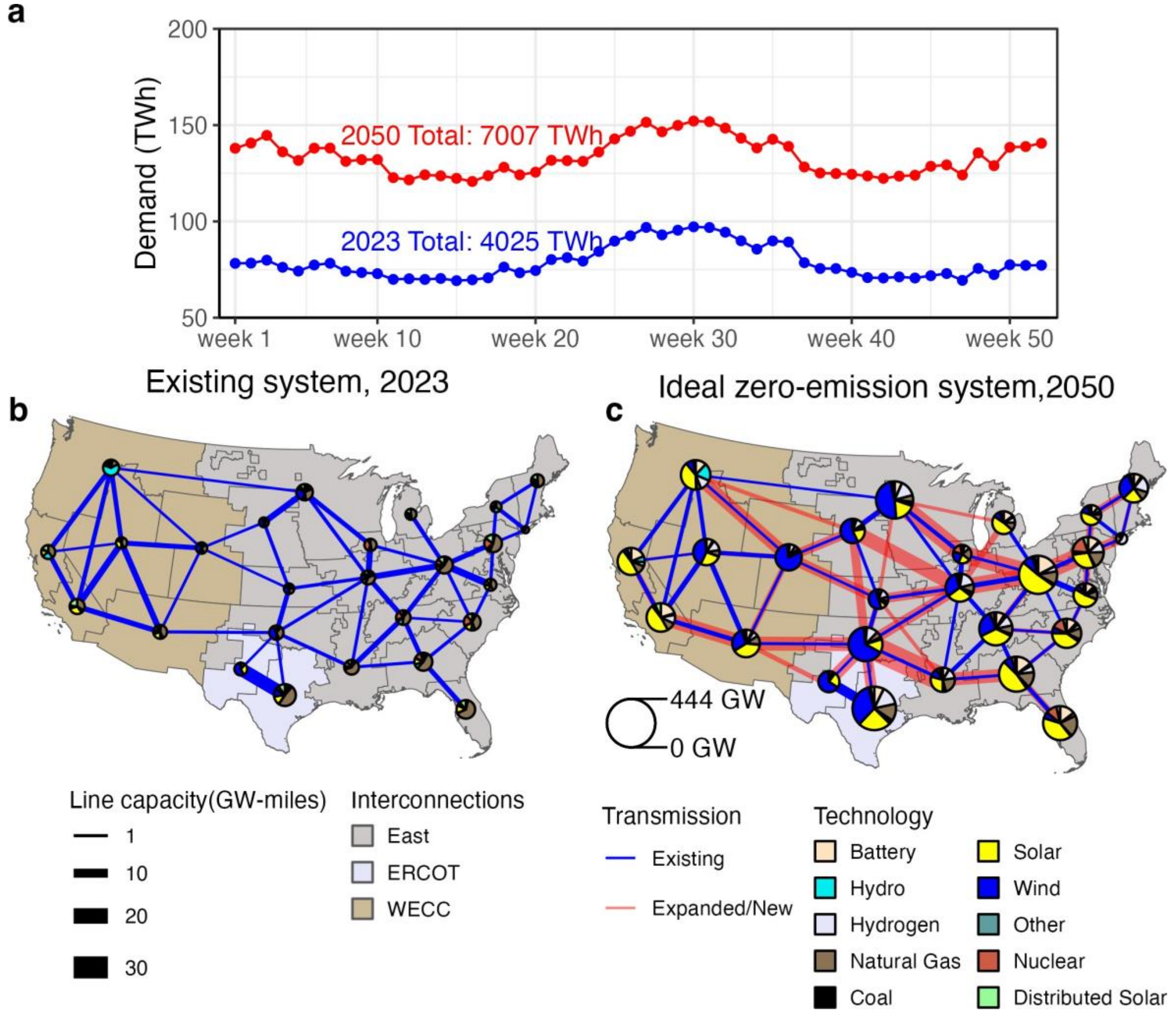

**Figure 1: Demand, generation, and transmission capacities in existing and idealized future zero-emission electricity systems.** Panel **a** shows weekly demand as modeled for 2022 and 2050. These demand data are derived from 2012 to synchronize with the weather data but are re-scaled to 2022 and projected to 2050 levels. Panel **b** shows 2022 generation capacities and inter-regional transmission. Note that total transmission capacity is less than 1 GW between the East and WECC. Panel **c** shows a least-cost, zero-emissions system for 2050 without superfluous constraints on the generation mix or transmission expansion.

The existing transmission system was not designed for the future, which will likely employ much more solar, wind, batteries, and other forms of storage, given their low and declining costs and lack of $CO_2$ and other pollution emissions. However, long-duration storage remains expensive, so there is an ongoing debate about what the future generation mix should look like (Jacobson et al., 2018; Pursiheimo, Holttinen and Koljonen, 2019; Qazi, 2022), and how the degree of interregional transmission expansion could factor into the generation mix (Staadecker et al., 2023), its geographic disposition, and overall cost. The future energy system is also policy-dependent. It will hinge on subsidies for clean energy, taxes or other restrictions on carbon dioxide and other pollutant emissions, local market rules and structures, and decisions by state and federal regulators.

To obtain a sense of what an idealized future system could look like, we used Switch (Fripp, 2012; Johnston et al., 2019; Staadecker et al., 2023; Hidalgo-Gonzalez, Johnston and Kammen, 2021; Sánchez-Pérez et al., 2022), a state-of-the-art, open-source power-system planning model, paired with PowerGenome (https://github.com/PowerGenome/PowerGenome), an open-source platform that compiles comprehensive weather and utility system data, to develop a least-cost plan for a zero-carbon power system for the continental United States in 2050. The model co-optimizes potential new generation capacities, storage, interregional transmission, and hourly chronological operation of the system for all 8,760 hours, which is essential for finding true least-cost high-renewable systems given their weather-driven variability and critical dependence on storage and transmission. Technology and fuel cost assumptions are derived mainly from the National Renewable Energy Laboratory's Annual Technology Baseline (National Renewable Energy Laboratory, 2022). This idealized system of the future, portrayed in panel **c** of Figure 1, is dominated by solar and wind generation, possesses ample storage in the form of batteries and green hydrogen, and increases interregional transmission capacity by over 200%. While optimal transmission expansion more than triples interregional transfer capacity—from approximately 0.13 TW to 0.62 TW—it simultaneously reduces the need for installed generation capacity by roughly 0.04 TW relative to scenarios with limited interregional transmission. This reduction reflects improved geographic smoothing of renewable output and lower reliance on redundant capacity across regions. Importantly, these changes occur in a broader context where wind, solar, battery, and hydrogen capacities expand by several hundred percent relative to existing levels. Since all large-scale infrastructure investments face siting, permitting, and political challenges, a more diversified expansion across transmission, generation, and storage may offer practical advantages over concentrating growth in any single category.

This idealized, zero-emissions system is 78% more costly per MWh than the least-direct-cost system that excludes any account of external $CO_2$ emissions costs. At a markedly lower cost (36% above least-cost, excluding emissions' costs), we could achieve roughly 95% decarbonization, a level of decarbonization consistent with the Environmental Protection Agency's estimated social cost of carbon dioxide of $190 per ton. The difference in cost and emissions between the carbon-priced and zero-emissions scenarios is mainly derived from natural gas with carbon capture and storage (CCS) replacing hydrogen. Note that these costs pertain only to generation and transmission and do not account for local distribution costs, which can comprise almost half of the retail electricity costs for residential customers today (Borenstein, Fowlie and Sallee, 2021). Hence, these decarbonization costs are a considerably smaller share of retail prices. However, these idealized systems optimize interregional

transmission, more than tripling it, including building new connections between the currently almost disjoint three interconnects (Eastern, Western, and ERCOT).

# 3 Scenarios considered

We consider a range of scenarios to evaluate the value of interregional transmission relative to potential substitutes. Each scenario jointly optimizes generation and storage capacities, transmission, and hourly chronological operation of the system for each of 52 weeks, subject to emissions and interregional transmission constraints of the given scenario. We use 1014 clusters of candidate renewable energy projects, corresponding to approximately 27 TW of potential capacity, with each cluster's variable capacity factors derived from weather data synchronized with demand. Both weather and demand are based on actual values from 2012; details are provided in methods.

## 3.1 Interregional Transmission and Emissions Constraints

To show how interregional transmission influences the cost of decarbonization, we built a series of scenarios with different restrictions on emissions and transmission expansion. These include three emissions scenarios: (i) a *least-direct-cost* system that ignores external costs of $CO_2$ and other pollution emissions; (ii) a *carbon-priced system* that assumes a $CO_2$ price of \$190 per ton; and (iii) a *zero-emissions system* with no $CO_2$ emissions. All three of these scenarios use the same projected demand for 2050, which is 74% greater than today's demand, roughly accounting for the growth of electric vehicles, partial electrification of heat, and some industrial activities. For each of these three emissions scenarios, we consider three interregional transmission scenarios: (a) *existing* transmission in 2022; (b) optimized expansion *within interconnects*, but no expansion between interconnects; and (c) *fully-optimized* expansion of transmission between regions both within and between interconnects. Crossing the emissions scenarios and transmission scenarios gives a total of nine scenarios labeled below.

**LE** Least-direct-cost, existing trans.
LI Least-direct-cost, within-interconnect trans.
**LO** Least-direct-cost, fully-optimal trans.
SE Carbon-priced, existing trans.
SI Carbon-priced, within-interconnect trans.
SO Carbon-priced, fully-optimal trans.
**ZE** Zero-emissions, existing trans.
ZI Zero-emissions, within-interconnect trans.
**ZO** Zero-emissions, fully-optimal trans.

In addition, we consider three supplemental scenarios:
BE Zero-emissions, existing trans, boosted demand.
BO Zero-emissions, fully-optimal trans, boosted demand.
ZE+ Zero-emissions, with transmission expansion limited to 25% of existing.

The BE and BO scenarios are like ZE and ZO except we increase demand for all hours by 25 percent on the three consecutive days that are most difficult to serve in each region in the ZE scenario. These scenarios test our conclusions against extreme conditions that are likely to favor transmission. This diagnostic may exaggerate the value of transmission because the difficult days are determined and boosted idiosyncratically, where in reality unusual events are likely to be correlated across regions, which would make transmission less valuable. The ZE+ scenario considers the value of increasing transmission modestly and sheds light on the most valuable expansions possible.

The main figures in this manuscript focus on **LE, LO, ZE,** and **ZO**, and we report results for the other scenarios in the supplement since they are similar to the cases shown here. In each case, all other aspects of the system are co-optimized, subject to transmission constraints. Existing and candidate generation technologies include solar, wind, batteries, hydrogen, natural gas with and without carbon capture, battery storage, nuclear, pumped-storage hydroelectric power, preexisting hydroelectricity, and distributed solar. Hydrogen production is produced endogenously via electrolysis and later used for generation via fuel cells, and thus serves as a form of long-term storage. The hydrogen supply chain is fully costed in the model, including capital, fixed, and variable operating costs for electrolyzers and fuel cells, as well as liquefaction and liquid hydrogen storage infrastructure. In our baseline assumptions, electrolyzer capital costs are approximately $770,000 per MW with a 40-year lifetime, while fuel cell capital costs are about $584,000 per MW with a 26-year lifetime.The model accounts for all system operations and maintenance costs, including those associated with existing transmission lines. By optimizing all options in conjunction with different transmission constraints, we see how portfolio adjustments compensate for less-than-ideal expansion of long-distance transmission.

### 3.2 Alternative Cost Scenarios

The scenarios described above adopt the "Mid" cost projections from NREL's 2022 Annual Technology Baseline (ATB). Given the considerable uncertainty surrounding future technology costs, and to better characterize how different resources serve as substitutes and/or complements to transmission, we conduct a sensitivity analysis along four key dimensions: (1) the capital cost of transmission expansion; (2) battery storage costs; (3) the cost of green hydrogen infrastructure (electrolyzers and associated O&M); and (4) the cost and availability of demand response. Each of these technologies provides system flexibility and may serve, at least partially, as a substitute for transmission, often in combination with changes in optimal solar and wind capacity. We focus this analysis on zero-emission scenarios, where the value of transmission is greatest, and explore a broad range of cost assumptions to capture the potential variability in technological development.

Specifically, we consider the following scenarios:

| | |
|---|---|
| **Transmission Cost:** | -50%, -30%, +30%, + 50% |
| **Battery Cost:** | -50%, -30%, +30% |
| **Hydrogen Cost:** | -50%, -30%, +30% |
| **Transmission & Battery:** | -30%, +30% |
| **Transmission & Hydrogen:** | -30%, +30% |
| **Demand Response:** | 10% of hourly load shiftable |
| **Adv. Conductors:** | -50% transmission cost; maximum 2X each existing link |

In the Demand Response scenario, up to 10% of the baseline demand in any hour can be shifted costlessly to another hour on the same day. To accommodate this shift, the load can increase up to 20% in any hour from baseline. But total demand in each day remains the same. This type of demand response is straightforward to solve because it involves a linear problem. In one sense, this kind of demand response is highly plausible in a clean energy future that will likely have a lot of demand from electric vehicles (included in our projections) and perhaps other devices that can be easily automated to charge or operate during the least cost times. Evidence already shows that electric vehicle owners are prone to adjust charging in this way (Bailey et al., 2025). However, this type of demand response does not capture a generalized response of demand to time-varying prices, which may change the overall level of use in addition to shifting. Consideration of generalized demand response makes the problem non-linear and requires iteration to solve, and is extremely costly on a national scale. We note that the values we calculate for demand response do not account for higher-frequency regulation services (shed service or shimmy service).

The advanced-conductor scenario (labeled $ZO^R$) considers a promising alternative: upgrading existing transmission corridors using advanced composite-core conductors and complementary grid-enhancing technologies. Chojkiewicz et al. (2024) find that advanced conductors carry approximately twice the power of conventional conductors for the vast majority of U.S. transmission segments–those under 50 miles in length, which account for roughly 98% of segments – at a total project cost well below that of new-build alternatives due to avoided right-of-way and structure costs. When this option is available, national models select roughly four times as much total interzonal capacity as under restricted greenfield scenarios, driven by lower per-unit costs attracting more total investment rather than by quadrupled per-line capacity. Consistent with the per-line empirical finding, our revised scenario assumes 50% below-reference transmission costs and limits expansion along each existing nodal link to at most *two* times current capacity. This represents a conservative approximation of what a comprehensive program combining advanced conductor replacement, dynamic line rating, and advanced power flow controllers could deliver on thermally-limited corridors; the INL Advanced Conductor Scan Report (2024) finds maximum ampacity improvements of up to 86% for thermally-limited lines without tower modifications, suggesting our two-fold assumption is at the upper end of what reconductoring alone can achieve on individual lines. We emphasize that this is an illustrative bounding case requiring more granular power-flow analysis to evaluate for specific corridors (Chojkiewicz et al., 2024).

# 4 Methods

The paper reports results from Switch, a structured power system planning model with input data compiled by PowerGenome. Both Switch and PowerGenome are open-source platforms with online documentation. We provide a brief characterization of the model here.

### Switch Characterization of the Planning Problem

Our analysis uses Switch (Fripp, 2012; Johnston et al., 2019), which is an open-source capacity expansion model that minimizes the net present value (NPV) of costs for an electricity grid across all investment periods and timepoints. It optimizes both the investment and operational costs subject to constraints such as emissions limits, emissions prices, or other policies, such as Renewable Energy Portfolio Standards (RPS).

Switch has a modular structure. Core modules define the time and spatial scale for the power system as well as an hourly load-balancing constraint and cost-minimizing objective function. Additional modules define physical components and add their costs to the objective function and their power contribution to the load balance. Other modules define constraints or costs to reflect policy choices, such as renewable portfolio standards or carbon caps. The modules we used in this study are described briefly below.

**Timescales**

Timescales defines the time horizon for the investment planning and energy balancing. Under the Switch modeling toolkit, the time resolution has a three-level hierarchy that accounts for the temporal dimension at various scales: periods, time series, and time points. Periods are a set of multi-year timescales that describe the times when the investment decisions are made. In this model, to focus squarely on the question of transmission, we assume just one investment period that stretches 10 years from 2041 to 2050 and refer to it as period 2050. The next level of granularity is the time series. This denotes blocks of consecutive time points within a period. An individual time series could represent a single day, a week, a month, or an entire year, or even a mix of blocks of different lengths. The length of time that energy may be stored is typically limited to the time series, which means that the amount of energy storage at the beginning of the time series is constrained to equal the ending amount at the end of the time series. (Hydrogen is an exception, as explained below.) There are 52 time series in this study, each one week long, with hourly time points from each week, comprising all 8760 hours in a year. Having 52 one-week time series instead of one time series of length 8760 saves considerably on computational cost while losing little practical precision since battery storage and most hydroelectric resources cannot be economically used for long-duration storage that exceeds a week.

**Financial**

The financial module defines the base year for the NPV calculation, the discount rate applied, and the interest rate used for financing capital investments. The base year in this study is year 2022 with the interest and discount rates set to 5% (real). The cost-minimizing objective function is defined in this module.

**Generation**

The generation modules define generation build-out options (both new and existing) and electricity dispatch, including fuel costs, variable O&M, and overnight build costs. Solar and wind installations have no fuel costs, but have variable capacity factors associated with each candidate project that account for hourly resource availability at that site. Switch has separate modules for storage, hydro, and endogenous hydrogen production and storage because of their unique operation and function. In specific, our model includes Li-ion battery storage—representing short-duration flexibility (typically under 8 hours) with a storage efficiency of 85%, pumped-storage hydroelectric power and endogenous hydrogen production which serves as a form of long-term storage. The storage module defines energy storage assets, optimizes new power and energy capacity, and optimizes their operation (charging and discharging). The hydro module enforces minimum and average flows for hydro resources for each time series. Newly built hydrogen generators can be implemented as generators that require the supply of hydrogen as an external fuel. In addition, Switch includes a module that produces hydrogen endogenously. In this module, Switch optimizes the amount of electrolyzer, fuel cell, liquefier, and hydrogen storage tank needed to be built and used for every model region. Storage tanks are sized to accommodate a whole year of hydrogen production and thereby facilitate seasonal storage, which is the predominant use.

### Transmission

The Transmission module represents the expansion and operation of the transmission assets using a transport model. In addition to optimized transmission expansion which allows additional capacity along any potential corridors, this module offers the option for users to disable expansion of any corridor using a binary parameter – trans_new_build_allowed. This flexibility allows us to consider scenarios without expansion or with partial expansion (e.g., within interconnects but not between).

### Policies

The policies subpackage has modules that enforce energy policy constraints such as RPS and carbon targets or carbon prices. The three decarbonization scenarios in our study are defined via the carbon policies module. (i) a least-direct-cost system that ignores external costs of CO2 and other pollution emissions; (ii) a carbon-priced system that assigns a cost of $190 per tonne of CO2 emitted; and (iii) a zero-emissions system with an emission cap of zero at 2050.

## PowerGenome and Principal Data Sources

One of the most difficult parts of running electricity capacity expansion models is assembling all the data. PowerGenome (Schivley, 2020) serves as a platform to generate input files for power system optimization models—including Switch, GenX, Temoa and USENSYS. The source data comes from a number of different sources, including EIA, NREL, and EPA.

### Model Regions

The extent of geographic coverage and number of regions is one of the first decisions to make when running an electricity planning model. Model regions in PowerGenome are derived from IPM regions. To align with the new-build resource cost multipliers for Electricity Market Module (EMM) regions in EIA's NEMS model, we group the IPM regions into 26 model regions in this study. These regions conjoin existing sub-regional balancing authorities. Region names are determined by matching the actual names listed in EIA's Open Data query search against each map and then looking at example API URLs. Data on existing generating units, cost estimates for new generating units, transmission constraints between regions, hourly load profiles and hourly generation profiles for wind & solar to construct the optimization problem are all parsed by these regions. The sources of these data are presented below.

### Existing and Candidate Generators

Existing generating units are from the latest version of form EIA-860 (U.S. Energy Information Administration, 2021), 2021 supplemented with 860m from June 2023. The cost and heat rates of new-build resources are provided by NREL's Annual Technology Baseline (ATB) 2022 (National Renewable Energy Laboratory, 2022).[1] Renewable resources are generally location-specific, with unique generation profiles and interconnection costs. Rather than representing all potential new-build renewable resources as individual sites, PowerGenome lets the user specify how much capacity of each resource type should be available for consideration in a model region and how many clusters the resource should be represented by. We use 942 clusters of candidate renewable energy projects, corresponding to approximately 21 TW of potential capacity. High-resolution weather data provided by Vibrant Clean Energy for 2012 (the year upon which our demand data is based) is used to construct hourly variable capacity factors for each cluster.

### Interregional Transmission lines

Existing transmission capacities between IPM regions are from EPA. Model regions in our study can consist of one or more individual IPM regions. When two or more IPM regions are combined to make a Switch model region, the transmission capacity between individual IPM regions is also combined. Transmission lines in this study are defined as connections between major metros in each region, with additional backbone networks connecting major metros within a region (if there is more than one). The cost assumptions for new transmission between the three asynchronous interconnections (Eastern, Western, and ERCOT) reflect back-to-back high-voltage direct current (HVDC) converter stations rather than long-distance HVDC lines, since the three interconnects are geographically adjacent at several points. The costs and line loss use transmission line segments for these converter stations are based on estimates from the transmission cost literature consistent with the approach described in Patankar et al. (2023), which uses a least-cost path method.

### Demand

Hourly demand starts with NREL EFS profiles National Renewable Energy Laboratory (2019). Stock values (historical) and hourly demand profiles of electrified end-use technologies like transportation, water heating, and space heating/cooling (derived from NREL EFS data) are subtracted from the 2019 EFS profiles. The remaining demand is inflated using sector-specific growth rates from EIA AEO 2022 (Nalley and LaRose, 2022). The future hourly demand from electrified end-use technologies are added back in using future stock values (from the REPEAT (Rapid Energy Policy Evaluation and Analysis Toolkit, 2022) scenario IRA_MID) and hourly demand profiles derived from EFS.

[1] Please note methane leakage from natural gas supply chains is not currently modeled in our analysis. Including upstream methane emissions at typical leakage rates (1–3% of produced gas) would modestly increase the effective $CO_2$-equivalent emissions of natural gas generation and CCS, making those technologies somewhat less attractive relative to zero-emissions alternatives. However, this omission does not affect our main findings since natural gas with CCS is only selected in our carbon-priced scenario (not in the zero-emissions scenarios that form the core of our analysis).

### Contingency Demand

In high-renewable settings, peak demand times are not typically the most difficult to serve, but rather *net* peaks, which equal demand less supply from intermittent renewables, solar and wind. In systems with a lot of storage, even the highest net peak days may be relatively easy to serve if adjacent hours or days have lower net peaks. To test the model's implications for transmission value in a robust way, we therefore found the highest three-day, demand-weighted average marginal cost in each region in the ZE scenario, and exogenously increased demand for all hours of these three days by 25%. Because transmission is heavily constrained in the ZE scenario, these three days tend to differ across regions, so boosting demand on these days ought to heavily favor transmission relative to baseline demand projections described above. We then resolve the zero-emissions scenarios, which we label BE and BO (for boosted demand with existing and optimal transmission, respectively). Results for these scenarios are reported in the supplement (Figure A8) and summarized in the main results. Within this system, the model predicts only a slightly higher value of optimal transmission expansion; it reduces overall costs by 7.7% instead of 7.0% relative to the existing transmission scenario.

The logic for why this diagnostic favors transmission is as follows: the ZE scenario already constrains transmission severely, so the most difficult days in each region are precisely those when local resources are most strained and when additional interregional transfer would be most valuable. Amplifying demand on those specific days therefore directly stresses the dimension of the system where transmission provides the greatest relief. Importantly, because the worst days are identified independently for each region, they do not correspond to a simultaneous continent-wide stress event (such as a widespread heat wave or winter storm), in which the correlation of stress across regions would reduce the effectiveness of interregional transmission. Our approach may therefore provide a slight upper bound on transmission value during single-region extreme events while underestimating it during correlated multi-region events, suggesting that the boosted-demand result of 7.7% is approximately representative of transmission value under stress.

## Limitations

Three aspects are important to discuss in the context of transmission line expansion. First, we model transmission capacity in an aggregated manner to represent important energy flows between load zones. The existing capacities we employ are consistent with those in EPA's Integrated Planning Model.[2] Our approach models the transfer capability of the electrical network but does not directly represent the physics of AC power flow. Hence, our findings should not be taken as a claim that there are few benefits associated with transmission expansion *within* individual regions in 2050. Our work sheds light on what are the most important electricity transfers between regions and interconnects to achieve zero emissions in 2050. Our findings speak to the modest economic impact of deploying "energy corridors" between interconnects rather than the value of specific transmission lines, for which power flow modeling would be required with an accurate model of the transmission network (e.g., California should be modeled with its approximately 9,000 buses).

[2] See https://www.epa.gov/power-sector-modeling/documentation-integrated-planning-model-ipm-base-case-v410. The regions in our model are aggregations of regions in the IPM model and transmission capacities between regions match aggregated capacities from their documentation.

Second, our study considers only a single weather year (2012) and coincident demand. We acknowledge this limitation and have worked to bound its consequences. The key distinction for interpreting our findings is between *level* questions and *gradient* questions. Studies that aim to identify the optimal system configuration – which technologies to build, in what quantities and locations – are sensitive to weather year choice because different meteorological conditions shift the global cost minimum. Our central question is different: we ask how much total system cost changes when the transmission constraint is relaxed, holding other substitutes available. This is a gradient or comparative statics question about the *marginal value of transmission relative to alternatives*, not the level of the optimum. When the cost surface is flat – as our results indicate, given the availability of many near-equivalent ways to balance the system – interannual meteorological variation may shift which specific technology mix is selected without materially changing the relative cost of transmission versus those alternatives. Our finding that one-fifth of the transmission benefit can be achieved with one-twelfth of the capacity, and that the cost difference between existing and optimal transmission is approximately 7%, reflects a system in which the optimizer has many degrees of freedom. This makes the gradient robust to perturbations, including changes in weather year, that shift the system's precise optimum but do not fundamentally alter the richness of the substitution possibilities. The contingency demand scenarios (BE, BO) test our conclusions specifically against amplified stress during the periods most favorable to transmission in the 2012 data. That the transmission benefit rises by less than one percentage point under this amplification (from 7.0% to 7.7%) provides direct evidence that the finding is not an artifact of a particularly benign weather year. We leave for future work the incorporation of multiple weather years, which would be especially valuable for refining investment recommendations at the regional level, where specific line choices are more sensitive to the particular distribution of weather and demand.

Third, we do not explicitly model reserves, resource adequacy requirements, or rare extreme weather events that fall outside the 2012 meteorological record. These considerations – which drive the reliability value of transmission analyzed from a different perspective by Botterud et al. (2025) – could lead us to underestimate transmission's contribution to system resilience. However, our focus is on the cost relationship between interregional transmission and its economic substitutes, and our boosted demand scenarios provide a targeted, if imperfect, test of robustness to more severe conditions than the baseline year.

## 5 Results

This paper uses average load-weighted, long-run marginal cost as the "wholesale cost" metric (the dual of the load-serving constraint in the optimization model). This corresponds to demand-weighted locational marginal prices (LMP) in wholesale power markets, assuming markets are perfectly competitive and in long-run equilibrium. The metric also includes the rents to scarce resources and can thus exceed cost metrics from engineering-based measures of levelized cost. However, we do not account for excess capacity payments, costs of obsolete or partially obsolete assets, within-region congestion costs, or any sub-optimal investment or operation costs of the system.

### 5.1 The Overall Value of Transmission Expansion

**Optimizing transmission for the least-direct-cost system reduces the average wholesale cost of power from 37.9 $/MWh under existing transmission to 37.6 $/MWh, a savings of just 0.8%** (Overall LE - Overall LO in Figure 2, panel **a**). This difference is barely visible in the graph due to its small size. Optimally expanding transmission only within interconnects reduces overall cost by 0.28 $/MWh, or almost all the savings from fully-optimal transmission. Although the least-direct-cost scenarios do not take into account emissions and assume inexpensive future natural gas following NREL-ATB projections, they still include a nearly five-fold expansion of solar and wind capacity relative to 2022 levels and increase electricity sector $CO_2$ emissions by around 10% relative to 2022. This emission increase follows from a 74% growth in overall demand and thus results in a 30% decrease in emissions intensity ($CO_2$ per MWh delivered) compared to 2022. Furthermore, the projected growth in demand embodies substantial growth of electric vehicles and electrification of heat, such that economy-wide emissions reductions would likely be much greater; a precise calculation is beyond the scope of this study.

**Optimizing transmission for the zero-emissions system reduces the average wholesale cost of power from 70.4 $/MWh under existing transmission to 65.5 $/MWh, a savings of 7.0%** (Overall ZE - Overall ZO in Figure 2). Optimally expanding transmission within interconnects reduces cost by 6.1% (not shown in the Figure). The relative savings of expanded transmission are thus much greater in the high-renewable zero-emissions system than in the least-direct-cost system. As shown in Figure 3, optimal transmission scenarios reduce wholesale marginal costs slightly throughout the distribution (also see Figure A3, in the supplement). If we limit transmission expansion to 25% more than existing ($ZE^{+}$), wholesale power costs equal 69.3 $/MWh, achieving about one-fifth of the benefits of optimal transmission with about 1/12 the expansion (Table 1). This exercise also highlights the most critical expansions, which we illustrate in the supplement (Figure A9). These expansions strengthen ties between ERCOT and the Eastern interconnect, the East and West interconnects, and Northeastern connections, especially around New York City.

| | Least-direct-cost | | | Zero-emissions | | | Carbon-priced | | | Supplemental | | |
|---|---|---|---|---|---|---|---|---|---|---|---|---|
| | **LE** | **LI** | **LO** | **ZE** | **ZI** | **ZO** | **SE** | **SI** | **SO** | **BE** | **BO** | **ZE+** |
| **Panel A: Cost components (Bil $)** | | | | | | | | | | | | |
| Fuels | 37.6 | 36.8 | 36.5 | 6.4 | 4.4 | 4.2 | 23.7 | 22.1 | 21.4 | 6.1 | 3.7 | 5.6 |
| OM | 4.1 | 4.0 | 4.0 | 2.0 | 1.4 | 1.4 | 6.9 | 6.8 | 6.5 | 1.9 | 1.2 | 1.8 |
| Hydrogen | 0.9 | 0.8 | 0.7 | 28.6 | 29.0 | 28.2 | 0.7 | 0.4 | 0.4 | 35.6 | 30.6 | 27.9 |
| Storages | 1.0 | 0.8 | 0.8 | 16.6 | 13.0 | 12.5 | 4.4 | 3.5 | 3.4 | 17.9 | 13.4 | 16.9 |
| Generation | 50.0 | 49.4 | 49.0 | 125.0 | 112.4 | 110.6 | 84.7 | 83.4 | 82.8 | 125.5 | 109.8 | 120.1 |
| Transmissions | 3.2 | 4.5 | 4.7 | 3.2 | 12.8 | 13.9 | 3.2 | 8.0 | 8.7 | 3.2 | 15.9 | 5.2 |
| SUBTOTAL | 96.8 | 96.3 | 95.9 | 181.9 | 173.0 | 170.7 | 123.6 | 124.1 | 123.2 | 190.2 | 174.7 | 177.5 |
| Emissions | 332.3 | 327.7 | 326.8 | 0.0 | 0.0 | 0.0 | 14.6 | 9.8 | 9.9 | 0.0 | 0.0 | 0.0 |
| TOTAL | 429.1 | 424.0 | 422.7 | 181.9 | 173.0 | 170.7 | 138.2 | 133.9 | 133.1 | 190.2 | 174.7 | 177.5 |
| **Panel B: Marginal cost ($/MWh)** | | | | | | | | | | | | |
| Load-weighted-MC | 37.9 | 37.6 | 37.6 | 70.4 | 66.1 | 65.5 | 54.8 | 51.5 | 51.1 | 71.7 | 66.2 | 69.3 |

**Table 1: Cost components and marginal cost for all scenarios. Panel A** reports the annualized costs in billions of 2022 dollars for each scenario, discounted at 5% (real) from the 2041-2050 investment period, to the present. Graphical displays of these numbers are in Figures 3 and A3. The *Least-direct-cost* scenarios minimize the net present value of all costs excluding emissions; the *Zero-emissions* scenarios force zero emissions; and the *Carbon-priced* scenarios optimize CO2 emissions assuming a social carbon cost of $190 per ton. Scenarios **LE, SE, ZE** restrict transmission capacities to their existing levels, but do account for maintenance of those lines; scenarios **LI, SI, ZI** optimize transmission capacities *within* existing interconnects, but do not disallow new transmission between interconnects; scenarios **LO, SO, ZO** fully optimize transmission within and between interconnects. Emissions costs multiply metric tons of emissions by \$190/t$CO_2$. **Panel B** shows the demand-weighted marginal cost, averaged over all hours and regions, in 2050.

## 5.2 The Costs and Benefits of Decarbonization

**The overall cost of decarbonization, defined as the difference between the average LMP of a zero-emissions system and a least-direct-cost system, is 27.9–32.5 $/MWh, or 74–86% of the least-direct-cost system, depending on the transmission scenario.** If emissions are valued at \$190/t$CO_2$, the avoided harm in the zero-carbon case is much greater than the extra direct costs in the power system. We find that full decarbonization has total societal costs—including both direct costs and emission impacts—that are 57%–59% below the least-direct-cost case. If societal costs are minimized instead of full decarbonization, direct costs increase by 22–28% (Table 1).

**The costs of decarbonization and benefits of optimal transmission vary across regions.** In Figure 2 panel **a**, regions are ordered from least to greatest cost in the scenario with zero emissions and existing transmission (ZE). These costs vary from about $50/MWh in TREW, which encompasses western Texas, to about $120/MWh in NYCM, which contains New York City and Long Island. It is clear from the map (panel **b**) that the transmission benefit for decarbonization varies across regions; in one case, expanding transmission increases the decarbonization cost (NYUP). The precise calculation for each point is the demand-weighted average hourly local marginal cost (LMP) for each region under each scenario. A neighboring region (NYCW), which encompasses New York City and Long Island, benefits the most from transmission expansion, reducing LMP by approximately 25%. Because the model optimizes the system's capital and chronological operation, and capital decisions are nearly continuous at the regional scale, the average marginal cost equals the average cost, including positive or negative economic rents to infra-marginal or sunk generation and transmission assets. We report on rents below and show how transmission changes their allocation, which may be relevant to the political economy of transmission expansion and of siting generation plants.

**To fully decarbonize using only existing transmission, investments in storage must increase by 32%, generation by 14%, and hydrogen electrolysis by 1.4% relative to the fully-optimized transmission scenario** (Figure 4, panel **a**). If transmission can expand optimally only within interconnects, generation and storage capacities change minimally relative to the fully optimized transmission scenario, while a small amount of hydrogen is used regardless of transmission. In the carbon-priced system, limiting to existing transmission requires storage and generation expenditure to increase by 29% and 22%, again with limited hydrogen use regardless of transmission (shown in Table 1 and supplement Figure A4). In general, the mix of generation is similar regardless of transmission but uses slightly more nuclear and solar and less wind when transmission is constrained (panels **c** and **d** of Figure 4). These compensating adjustments in generation and storage expansions are somewhat more expensive than the optimized transmission expansions that they replace (Figure 4, panel **a**). The adjustments are modest, however, relative to the eight- to nine-fold increase in solar and wind (from the end of 2023), among other capacity increases, needed for decarbonization.

Generation capacities and their geographic locations, including those for solar, wind, batteries, hydrogen, nuclear, and other sources, are generally similar regardless of transmission expansion and vary mainly with the degree of decarbonization. We illustrate these differences in Figure 5, which plots the capacities in each region and type when only existing transmission is used against the capacities when interregional transmission is fully optimized. The similarity of these generation mixes has a practical implication for planning: it suggests that generation expansion should proceed similarly regardless of inter-regional transmission expansion, especially during the early to intermediate stages of decarbonization. In the supplement Figure A5, we show similar comparisons for the carbon-priced scenarios that assume a carbon price of $190/ton.

In addition to the technologies shown, the model includes options for carbon capture and storage (CCS) paired with natural gas and coal generation facilities. Because CCS is not zero-emissions, it is only selected in the carbon-priced scenarios, with details provided in the supplement (Figure A4).

When we make the most costly days in each region even more difficult to serve by boosting demand 25 percent – scenarios **BE, BO** – we find the value of transmission increases from 7.0% to 7.7%, which indicates our findings are unlikely to be fundamentally altered by limited weather and demand outcomes (Figure A8 and Table 1).

### 5.3 How Demand Response and Infrastructure Costs Affect the Value of Transmission

To explore the influence of demand response and cost assumptions on transmission value, we focus on the zero-emissions scenarios where transmission has the greatest potential benefit. The specific scenarios considered are outlined in Section 3, and results are summarized in Table 2 and Figures 8, 9, and 10. We summarize these results below.

**Demand response and critical infrastructure costs can significantly affect the total cost of a decarbonized system and influence generation mix and siting, but generally have a modest effect on the net benefits of optimal transmission expansion.** Specifically, demand response (10% within-day shiftable load) reduces the total cost of a zero-emissions system by \$3.3/MWh under optimal transmission expansion and by \$3.4/MWh under existing transmission. Thus, this modest level of demand response yields cost savings equal to roughly two-thirds of the benefit of moving from existing to optimal transmission, or about 5% of total system cost. The difference in savings between the ZE and ZO scenarios implies that demand response reduces the value of transmission expansion by only \$0.1/MWh (Figure 8 and Table 2). The demand response savings reported in this study assume costless shifting of demand within each day, and should therefore be interpreted as an upper-bound benchmark of the potential system value of within-day demand flexibility under idealized conditions. Note, however, that Imelda, Fripp and Roberts (2024) find that generalized price responsiveness would yield greater benefits than within-day shifting, as it can address broader energy shortages and surpluses.[3]

**A 50% reduction in battery costs from the baseline lowers system-wide costs by \$4.8 to \$5.8/MWh, comparable to the savings from moving from existing to optimal transmission, holding all else constant.** The resulting \$1/MWh difference indicates that substantially lower battery costs slightly reduce the marginal value of interregional transmission expansion. A smaller 30% battery cost reduction decreases system-wide costs by \$2.2 to \$2.7/MWh, whereas a 30% increase raises costs by \$1.7 to \$1.9/MWh. Overall, battery cost reductions benefit the system more than equivalent cost increases harm it, and the effect on transmission value remains modest.

Green hydrogen cost reductions yield more surprising results. In the ZO scenario, a 50% cost reduction produces greater system-wide savings (\$6.1/MWh) than an equivalent battery cost reduction. In the ZE scenario, by contrast, reduced battery costs (\$5.8/MWh) are slightly more beneficial than reduced hydrogen costs (\$5.6/MWh). Notably, lower hydrogen costs slightly *increase* the value of transmission, in contrast to lower battery costs, which tend to reduce it.

Differences in the resulting generation mix help explain why hydrogen complements transmission expansion while battery storage substitutes for it. Lower battery costs lead the model to favor more solar capacity, less wind, and reduced transmission investment. In contrast, lower hydrogen costs shift the mix toward more wind, less solar, and greater transmission investment. In practice, the model tends to locate electrolyzers in wind-rich regions, where low marginal-cost generation and surplus wind production make hydrogen production economically attractive. Hydrogen storage and reconversion (e.g., via fuel cells) may occur either near these resource regions or closer to load centers, depending on transmission constraints and cost trade-offs. Batteries pair naturally with solar generation due to the predictable, short-term day–night variability of solar output. Moreover, solar resources are widely available and less land-constrained, making them less dependent on interregional transmission. By contrast, hydrogen supports wind: unlike batteries, hydrogen can mitigate long-duration imbalances and broader energy shortages, not just short-term fluctuations—challenges that are particularly relevant for geographically concentrated wind resources. Because wind benefits from expanded transmission, and lower hydrogen costs favor wind deployment, reductions in hydrogen costs indirectly increase the value of transmission.

[3] Generalized demand response refers to consumption sensitivity to price that involves substitution across hours and changes in total quantity consumed, not just shifting demand within the day. Modeling generalized demand response is computationally intensive, especially at a national scale, because it makes the problem non-linear and requires iteration. We leave this for future research.

| Scenario | Transmission Scenario: Existing (ZE) ($/MWh) | Transmission Scenario: Optimal (ZO) ($/MWh) | Difference: Δ ($/MWh) | Difference: Δ (% of ZE) |
|---|---|---|---|---|
| Baseline | 70.4 | 65.5 | 4.9 | 6.9 |
| Demand Response | 66.8 | 62.2 | 7.8 | 7.0 |
| Battery Cost | | | | |
| + 30% | 72.3 | 67.2 | 5.1 | 7.1 |
| − 30% | 67.7 | 63.3 | 4.4 | 6.4 |
| − 50% | 64.6 | 60.7 | 3.9 | 6.2 |
| Hydrogen Cost | | | | |
| + 30% | 72.7 | 68.4 | 4.3 | 5.8 |
| − 30% | 67.5 | 62.2 | 5.3 | 7.8 |
| − 50% | 64.8 | 59.4 | 5.4 | 8.3 |
| Transmission Cost | | | | |
| + 50% | 70.4 | 67.2 | 3.2 | 4.5 |
| + 30% | 70.4 | 66.6 | 3.8 | 5.3 |
| − 30% | 70.4 | 64.3 | 6.1 | 8.6 |
| − 50% | 70.4 | 62.7 | 7.7 | 10.9 |
| − 50% Advanced conductors (2× cap) | 70.4 | 69.3 | 1.1 | 1.6 |
| Transmission & Battery Cost | | | | |
| + 30% | 72.3 | 68.4 | 6.0 | 5.4 |
| − 30% | 67.7 | 61.7 | 6.0 | 8.8 |
| Transmission & Hydrogen Cost | | | | |
| + 30% | 72.7 | 69.4 | 3.3 | 4.5 |
| − 30% | 67.5 | 60.4 | 7.1 | 10.5 |

**Table 2: The value of transmission expansion in a zero-emissions system under demand response and alternative cost assumptions.** The table reports how overall costs (per MWh) and the value of transmission change under alternative assumptions about demand response and critical capital costs. The baseline refers to the difference between ZE and ZO under NREL "Mid" cost assumptions and is shown in other figures and tables. As battery costs, transmission costs, or hydrogen costs vary from this baseline, the overall $/MWh cost of decarbonization and transmission value can increase or decrease, holding all else constant. Note that under the alternative transmission scenarios, we only vary the cost of capital expansion; the costs of operations and maintenance are fixed at baseline levels, including maintenance of existing lines. Under Transmission, the “-50% Advanced conductors (2× cap)" scenario disallows new connections between regions and limits expansion along any nodal link to two times the current capacity.

As expected, lower transmission costs increase the value of expansion, while higher costs reduce it. However, even when unconstrained, **50% below-reference transmission costs reduce the total cost of a zero-emissions system ($2.8/MWh) less than 50% below-reference battery or hydrogen costs**. This gain is achieved in part through an eight-fold increase in interregional transmission capacity

(Figure 10).

Although technology-related cost declines are less dramatic for transmission, non-physical factors—such as transaction costs and environmental impacts of securing new rights-of-way—are likely critical. Grid-enhancing technologies, including dynamic line rating and advanced power flow controllers, can increase capacity on existing lines at relatively low cost.

The advanced-conductors scenario ($\mathrm{ZO}^R$), with 50% below-reference transmission costs but limiting each nodal link to a two-fold capacity increase and disallowing new connections between regions, yields more limited system-wide benefits than the unconstrained cost-reduction scenarios. **$\mathrm{ZO}^R$ reduces system costs by only \$1.1/MWh (1.6%) relative to ZE—far below the \$7.7/MWh (10.9%) achievable under unconstrained 50%-below-reference transmission costs, and below comparable reductions in battery or hydrogen costs (\$4.8–6.1/MWh)** (see Figure 10). The gap of \$6.6/MWh between $\mathrm{ZO}^R$ and the unconstrained case is larger than the entire baseline optimization benefit (\$4.9/MWh), indicating that the per-link cap and the prohibition on new connections are binding constraints.

A telling comparison: $\mathrm{ZO}^R$ achieves exactly the same system cost (69.3 \$/MWh) as $\mathrm{ZE}^+$, which permits only 25% expansion above existing capacity but allows new connections on the highest-value corridors. $\mathrm{ZO}^R$ carries nearly twice as much total capacity as $\mathrm{ZE}^+$ at half the unit cost, yet the two scenarios cost the same. The $\mathrm{ZO}^R$ scenario is constrained in two ways simultaneously: the 2× per-link cap prevents substantial deepening of capacity on existing high-value corridors, and the prohibition on new connections blocks the new linkages that $\mathrm{ZE}^+$ identifies as especially valuable. Either relaxation alone—more capacity depth on existing lines, or selective new connections—would likely improve results, and the combination would likely be more valuable still. The 2× limit reflects what reconductoring alone can achieve on thermally-limited individual lines; unlocking greater system benefit would require either a more aggressive conductor program that approaches the system-level 4× outcome documented by Chojkiewicz et al., or pairing upgrades with targeted new construction on key corridors. Advanced conductors deployed within existing rights-of-way cannot substitute for these new connections, and the optimizer compensates largely by building more solar and battery storage, replicating much of the same substitution pattern observed under hard transmission constraints in ZE. Advanced conductors may nonetheless have real potential, particularly if combined with selective new-build connections on key corridors or complemented by lower battery costs.

Finally, less tangible costs from permitting delays and land acquisition could push transmission costs well above baseline. The implications of above-baseline transmission costs (+30% and +50%), as well as combined scenarios with transmission +/−30% costs and +/−30% battery or hydrogen costs, are shown in Table 2 and Figures 8, 9, and 10. These scenarios reduce transmission value relative to the baseline and shift the generation mix toward solar and batteries and away from wind. The mix and value of transmission are less sensitive to cost increases than to equivalent cost declines.

### 5.4 The Effects Decarbonization and Transmission Expansion on Price Variability

**Transmission expansion has little influence on price variability, while the extent of decarbonization significantly increases price variability** (Figure 3). In competitive markets, wholesale prices equal marginal cost, the distributions of which we can infer from the model's dual variables associated with each hour's balancing constraint. We show these distributions as bar plots with binned ranges

because of the significant mass point at zero and to ease comparison of scenarios with and without transmission expansion. The supplement shows cumulative density functions (Figure 2). We discuss possible real-world departures from competitive pricing below.

The marginal-cost distributions for the least-direct-cost scenarios have about 90 percent of MWh between 20 and 40 dollars per MWh. About 2 percent of MWh have a marginal cost of zero or nearly so (< \$1/MWh), and about 4 percent have marginal costs above 40 dollars per MWh. The zero marginal-cost times occur during curtailment events (wind and/or solar energy are discarded).

In contrast, the zero-emissions scenarios have starkly greater variability, with 40-45 percent of delivered MWh having a zero or near-zero marginal cost (< \$1/MWh). The large share of "free" energy occurs because solar and wind capacities are optimally built out to help serve demand on days with low to medium wind and sunlight and high demand. These times are seasonal, tending to occur in late fall and winter. As a result, there are often surpluses during other seasons, especially spring and summer, when wind and sunlight are more abundant, leading to many stretches of time with zero marginal cost. During these times, renewables are curtailed because supply exceeds demand and batteries are full. There can be zero-cost times without simultaneous curtailment if there is more energy in batteries than could be depleted before being fully charged again with excess wind and solar.

In zero-emissions scenarios, the spread of marginal cost above near-zero is lumpy, with just 10 percent of MWh priced between \$1 and \$90, 20-30 percent spread evenly between \$90 and \$100 per MWh, 20-25 percent spread between \$100 and \$200 per MWh. Marginal costs above \$200 are rare and extremely rare above \$1000/MWh. Transmission mainly increases the frequency of hours in the near zero and \$100 to \$200 ranges and decreases them in the \$90 to \$100 range. Again, the influence of transmission is modest.

In the carbon-priced scenarios, the distribution looks like a mixture of the least-cost and zero-emissions scenarios, with about 20 percent of MWh priced near zero, about 50 percent in the \$20-50 range, and 25 percent in the \$80 to \$200 range.

In all three cases, the influence of transmission on the distribution of marginal costs is modest. The extreme variability of marginal cost in higher-renewable scenarios, especially the high frequency of zero-marginal-cost energy, suggests that flexible demand could be valuable in ways quite unlike conventional systems where only critical peaks matter (Imelda, Fripp and Roberts, 2024).

**Hourly operational detail in ERCOT shows how transmission expansion can aid resiliency during the most constrained times** (Figure 7). The model optimizes hourly operations and interregional flows in each region, revealing how the system achieves balance. Here, we consider a costly day in a zero-emissions scenario in the most isolated region, TRE, the eastern part of ERCOT, which has most of the interconnect's generation and demand. In two of the four scenarios (LO, ZE), the day depicted (December 13$^{th}$, 2050, based on actual weather data in 2012), is the most expensive in the region. This is a cold day in late Fall near the solstice, when solar resources would generally be more limited, and with unusually little wind.

Compared with the least-cost scenarios, decarbonization is achieved via substantially greater capacities of solar and wind paired with significant use of batteries and hydrogen, which replaces natural gas and some old-vintage coal. The deficit from renewable energy is mostly counterbalanced with natural gas in the least-direct-cost scenarios and with imports, battery storage, and hydrogen in the zero-emissions scenarios. Notably, even the least-direct-cost scenarios use batteries to help serve the net peak at 7 pm, when wind and solar power fall to near-zero.

The effects of optimizing transmission are subtle and difficult to discern in the least-direct-cost scenarios. In these scenarios, some power is imported from the Western part of ERCOT (TREW) at 1 pm and 3 pm of December 13$^{th}$, 2050 when transmission is fixed (panel **a**), but less is imported under optimal transmission, because, in this case, more natural gas capacity is built so the region can export more on other days. The generation mix is largely the same when transmission is fully optimized (panel **b**), except there is slightly more gas generation and less battery use during the evening peak. The peak hour marginal cost falls from about $15,000 to about $6,000 when optimizing transmission.

The zero-emissions scenarios (**c** and **d**) show a somewhat greater influence from optimizing transmission. Generation mixes are similar regardless of transmission, but there is noticeably more battery use and slightly less wind in the scenario with existing transmission. And while imports help, especially in the optimal transmission scenario, it is interesting that imports occur midday to help replenish batteries, and much less during the expensive evening peak. The peak marginal cost of about $35,000 per MWh falls to about $20,000 with expanded transmission. And where the nighttime hours after the peak stay elevated at $5,000 per MWh with existing transmission, marginal cost falls back to very low levels when transmission is optimally expanded.

### 5.5 The Influence of Transmission Expansion on Economic Rents

**Transmission expansion increases economic rents for some resource owners while reducing them for others** (Figure 6). If resource owners are paid a price equal to the incremental cost of power in each region and hour, then all variable and fixed costs are recovered in long-run equilibrium, which is what the model assumes. In addition, some owners of scarce resources glean rent above their total costs. The potential rents identified in this study pertain to heterogeneous solar and wind resources, as well as potentially constrained transmission resources. Areas with a comparative advantage in clean resources, such as unusually large or well-timed amounts of solar radiation or wind, earn rent because they provide more inframarginal value than marginal resources do. These rents are logically highest in the zero-emissions scenarios.

Transmission expansion increases rents for some resources and regions, while decreasing them for others. For example, a region with high-quality wind or solar resources benefits from expanded transmission and trade with regions that have lower-quality wind and solar resources, while a region with lower-quality wind and solar resources may lose out when enhanced transmission instills greater competition from neighboring areas.

Combining rent and marginal cost changes, which mostly benefit consumers, all regions gain from optimal transmission expansion. However, since some resource owners lose rents relative to the no-expansion case, diverging stakeholder interests could hamper efforts to expand transmission. Figure 6 illustrates these divergent interests. The graph shows gains and losses in solar, wind, nuclear, and transmission rent when going from the ZE scenario to the ZO scenario. Some rent gains are intuitive, such as wind in SPSS, the Rocky Mountain regions, and MISW, and rent losses in other, less windy areas. Gains in solar rents in NWPP and PJMW and loss of solar rents in CASO may be more surprising, but less so if one considers time patterns of net demand in these regions relative to neighbors. MISS loses rents for all three resources due to greater imports, with much greater benefits from transmission going to consumers (Figure 2). Nuclear plants generally lose rent from greater transmission, as it brings greater competition from solar and wind and much less is selected for expansion. (Rent accrues to existing nuclear if not expanded, as capital costs for existing are assumed sunk.) Existing transmission owners also lose substantial rents, as pointed out by Hausman (2024). The losses considered here, however, are forward-looking, and account for emerging substitutes and substantial projected demand growth.

It is interesting to compare the rent implications of transmission expansion under full decarbonization versus carbon-priced optimization, reported in the supplement (Figure A6). The geographic allocations are similar but tend to be smaller in magnitude, especially for nuclear. Transmission rent losses are more concentrated in the Northeastern regions.

## 6 Conclusion

While the optimal expansion of interregional transmission would more than quadruple current capacity—a result consistent with other studies—we find that the net benefits of this expansion are modest relative to the overall cost of a decarbonized system. Substitutes for transmission include battery storage, additional renewable generation, and nuclear expansion. Transmission may improve resiliency in some locations, but the distribution of marginal costs (i.e., competitive wholesale prices) remains broadly similar regardless of interregional expansion. Consistent with Wongel and Caldeira (2023), our findings indicate that there are many ways to firm solar and wind power, even with a more granular representation of resources and explicit treatment of transmission.

The study also clarifies how renewable resources substitute for or complement each other and interregional transmission. Transmission constraints increase reliance on solar and batteries while reducing optimal wind and hydrogen capacity; conversely, greater transmission favors wind and hydrogen.

We find moderate diminishing returns: about one-fifth of the benefits of full expansion can be achieved with just one-twelfth of the capacity, with the most valuable additions linking the three interconnection regions and the New York City area with neighbors. Optimal interregional transmission is highly sensitive to costs, reaching nearly eight times today's capacity—more than double our baseline—if costs are halved, but falling by about one-quarter if costs are 50% higher. Our advanced-conductor scenario ($ZO^R$)—which assumes half-costs but caps each link at two times current capacity and disallows new connections, consistent with the per-line finding of Chojkiewicz et al. (2024)—results in 2.4 times current capacity but only 1.6% cost savings relative to ZE. This is well below the 7% savings from unconstrained optimization and below comparable reductions in battery or hydrogen costs. The scenario costs \$6.6/MWh more than unconstrained 50%-below-reference transmission—a gap larger than the baseline optimization benefit itself—and achieves the same system cost as $ZE^+$, which permits only modest expansion but allows new connections. This comparison reveals that the value of transmission is concentrated in new linkages, not in cheapening existing corridors. Advanced conductors may nonetheless have real potential if paired with selective new-build connections on key corridors or complemented by lower battery costs; more granular power-flow modeling would be needed to identify where within-right-of-way upgrades provide the greatest system value..

Interregional Transmission also affects rents for inframarginal wind, solar, and nuclear resources, which may illuminate political-economic forces shaping support or resistance to expansion and to substitutes or complements like solar, batteries, wind, and nuclear. The transmission rent analysis is illustrative of a general principle — any significant capacity addition or constraint reshapes the distribution of rents — and we focus on transmission specifically because it is the subject of our study and because its geographic specificity creates particularly clear distributional patterns. We note that a parallel analysis of, for example, large-scale battery deployment would similarly affect solar and wind rents in ways that could shape political coalitions for or against battery expansion. Our use of average marginal cost facilitates these rent calculations and underscores the much greater variability in wholesale prices under high-renewable futures. We also disallow flexible ramping of nuclear. Because nuclear is costly but slightly cost-reducing in decarbonized systems, inflexibility raises the value of transmission. While new plants may be more flexible than we assume, we restrict flexibility for consistency and to err on the side of assigning more value to transmission. This and other assumptions likely overstate the benefits of expansion, which makes it important to note key qualifications.

First, our advanced-conductor scenario provides an illustrative bounding case but does not model the full range of grid-enhancing technologies (GETs) that could expand interregional capacity at low cost.[4] Such technologies include dynamic line rating and transformer upgrades. Dynamic line rating is particularly promising in high-renewable systems, as difficult-to-serve hours occur in cooler months when wind output is higher, allowing greater safe transfer capacity than conventional static ratings.[5]

Second, we assume perfectly competitive markets. In practice, congestion may create opportunities for market power, allowing generators or transmission owners to withhold capacity and raise prices (Borenstein, Bushnell and Wolak, 2002). Transmission could thus promote competition, but those benefiting from scarcity or market-power rents may resist expansion. If reducing market power is a priority, much more expansion may be justified than our efficiency-based optima suggest.

Third, our demand response assumptions are limited. Many forms of demand adjustment – EV charging and vehicle-to-grid services, building insulation, thermal storage and smart HVAC/water heating, thermostat adjustments, and industrial demand shifting – could substitute for interregional transmission, storage, and generation. Some of these responses are already embedded in demand data, but high-renewable stress periods occur at different times than in the baseline, so existing demand flexibility could reduce system costs and the value of transmission more than we estimate.

Fourth, growing variability in marginal costs creates opportunities for utilities, communities, and intermediaries to arbitrage wholesale price differentials. If such demand-side strategies proliferate, the value of interregional transmission could be well below our estimates. Future work should examine interactions between transmission, demand response, and distributed resources across scales. A robust, price-responsive demand side would also mitigate market power during congestion.

Finally, substitutability depends on the timely availability of alternatives. Interconnection queues for renewables and storage remain a major barrier: at the end of 2023, over 2.5 TW of capacity awaited grid connection—enough to decarbonize the projected 2050 electricity sector with 75% higher load than in 2022—yet wait times are long and many projects never reach completion (Rand et al., 2023). Our study does not address interconnection bottlenecks or full power-flow dynamics but does include approximate spur-line and upgrade costs to connect new resources. This distinction is important: while local transmission upgrades are essential, they should not be conflated with large-scale interregional expansion.

[4]See Chojkiewicz et al. (2024) and https://www.energy.gov/oe/grid-enhancing-technologies-improve-existing-power-lines.

[5]See, for example, https://www.nrel.gov/grid/transmission-series.

# Figures

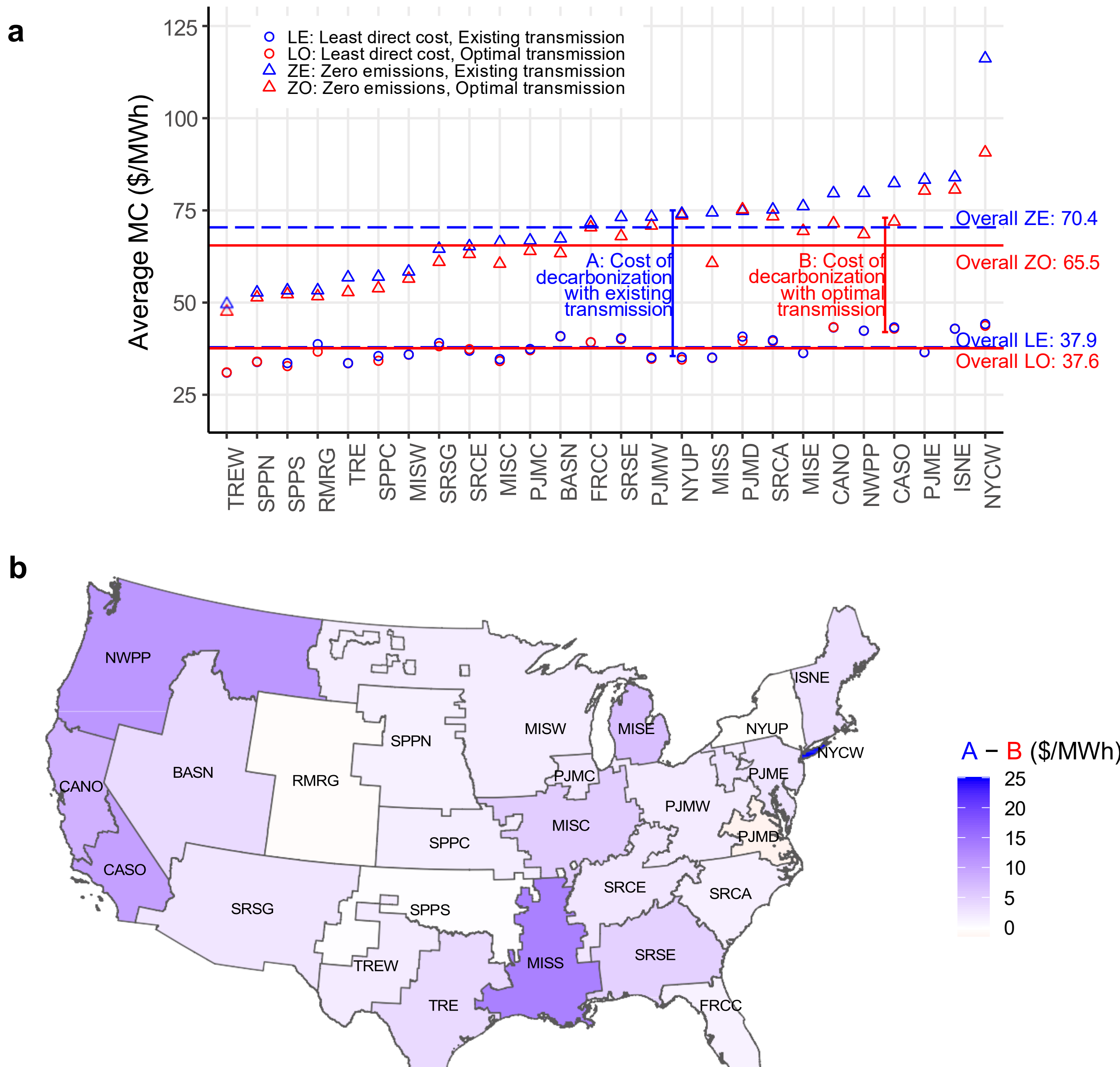

**Figure 2: Cost per MWh for the different emissions and transmission scenarios.** Panel **a** shows the demand-weighted average marginal cost for each region in 2050 in four scenarios, least-direct-cost (triangles) and zero-emissions scenarios (circles), each with existing (blue) and optimized (red) transmission. Comparing circles to triangles of the same color gives the region's cost of decarbonization, with blue indicating the cost without transmission expansion (difference A) and red indicating the cost with optimized transmission expansion (difference B). Comparing the same shapes of different colors gives the net savings from expanded transmission. Panel **b** shows a map of the difference in differences (A-B): the cost savings ($/MWh) from fully decarbonizing using optimal transmission instead of using only existing transmission.

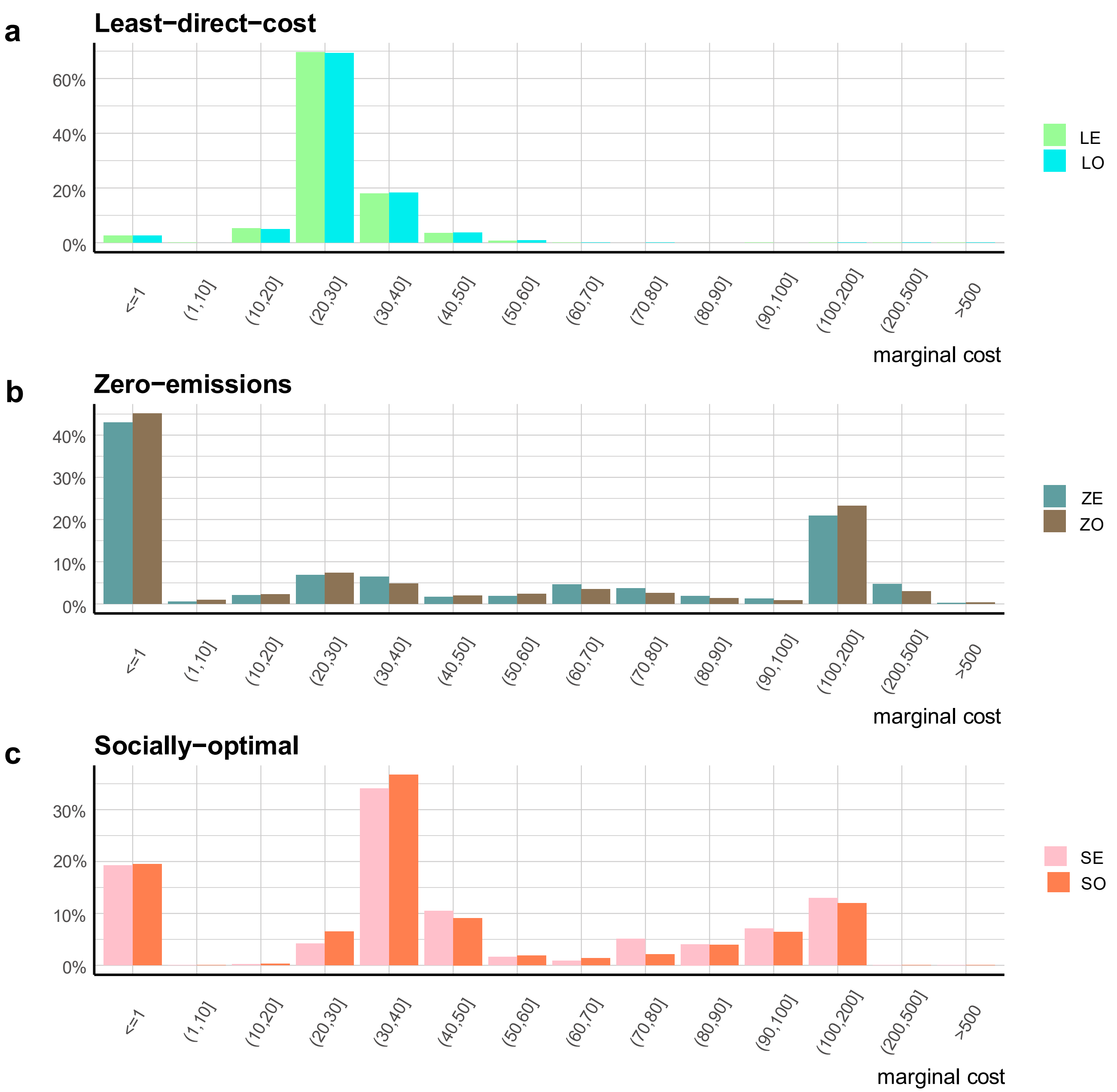

**Figure 3: Count percentage of marginal cost.** This graph shows the count percentage for hourly marginal cost across all MWh in all continental U.S. regions in 2050. Six scenarios are depicted: Panel **a** shows the system of least-direct-cost, with existing and optimal transmission (LE & LO); Panel **b** shows the system of zero-emissions with existing and optimal transmission (ZE & ZO); Panel **c** shows the system of carbon-priced, with an assumed price of $CO_2$ emissions of $190 per ton, which achieves roughly 89% reduction of emissions from the electricity sector relative to 2022.

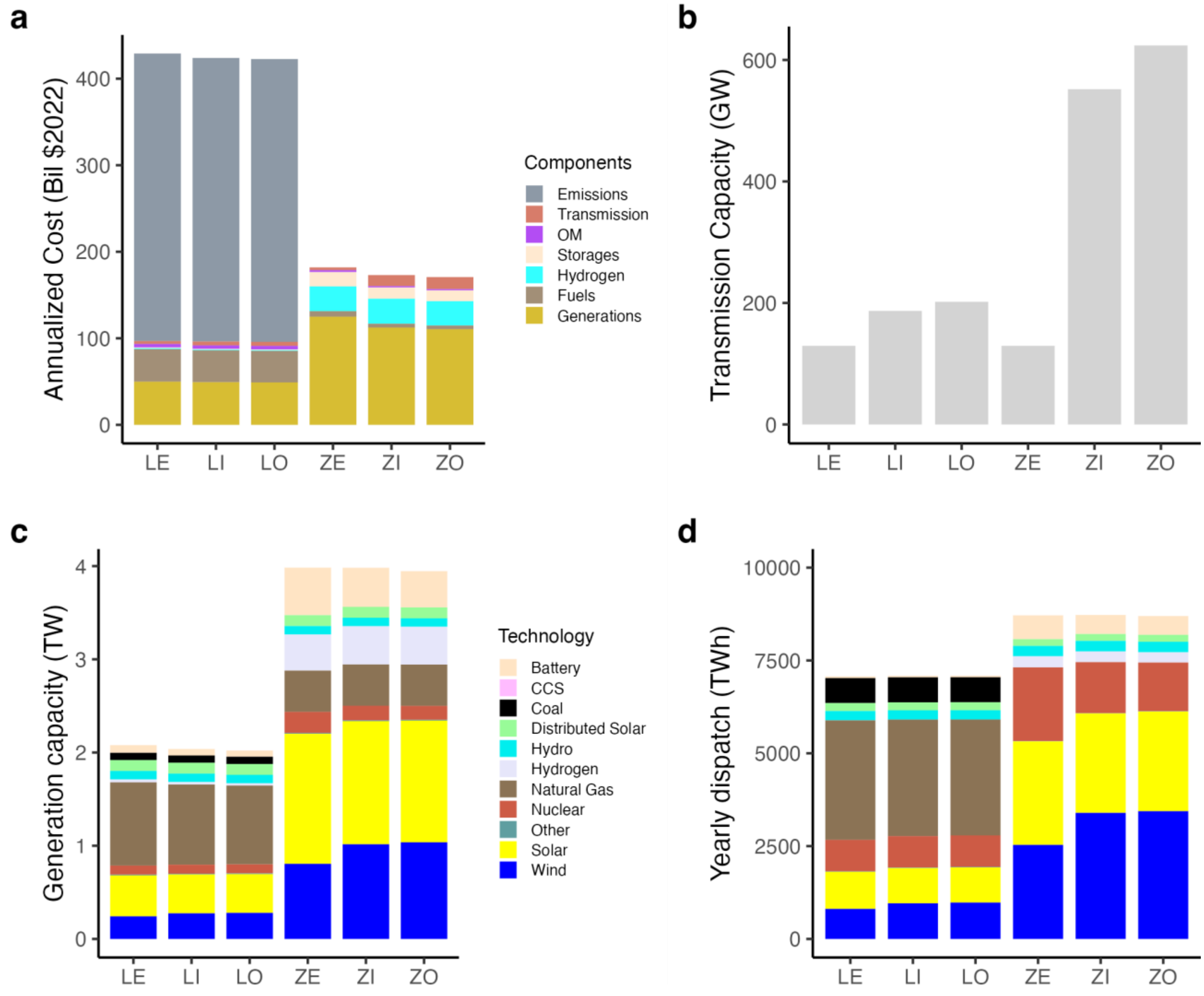

**Figure 4: Comparing component costs and capacities.** The graphs compare costs, generation mixes, and transmission capacities across six scenarios, LE (least-direct-cost with existing transmission), LI (least-direct-cost with optimal within-interconnect transmission), LO (least-direct-cost with optimal fully-optimized transmission), ZE (zero-emissions with existing transmission), ZI (zero-emissions with optimal within-interconnect transmission), and ZO (zero-emissions with optimal fully-optimized transmission). Panel **a** shows broadly categorized cost components; panel **b** shows total transmission capacity in each scenario (GW); panel **c** shows total generation capacities (TW) in each scenario; and panel **d** shows the share of dispatch (source of energy consumed) in each scenario.

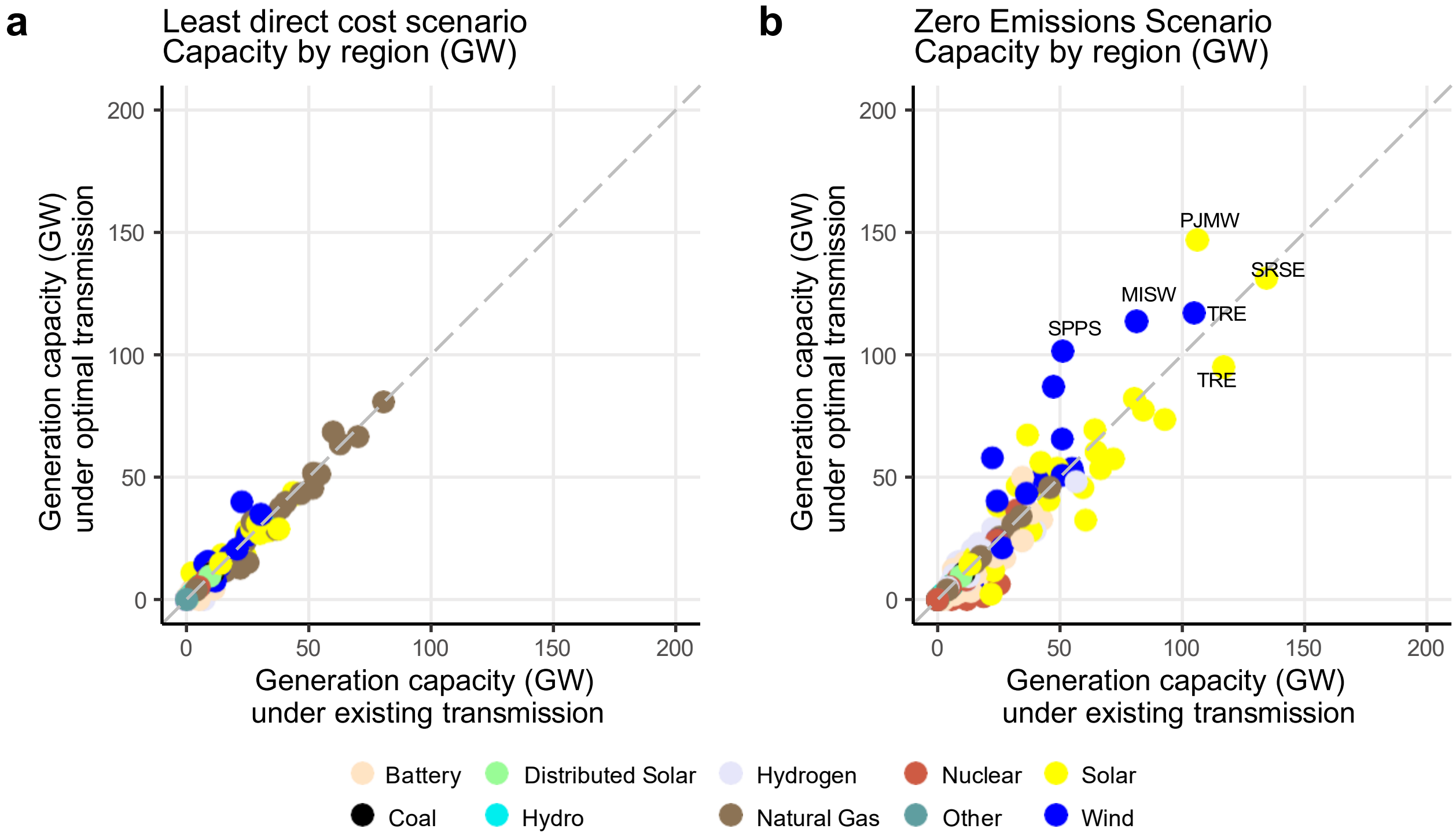

**Figure 5: Comparison of regional capacities across transmission scenarios.** These scatter plots show how transmission expansion influences the mix of generation capacities across regions. It plots generation capacities in each region and type when only existing transmission is used against capacities when interregional transmission is fully optimized. Panel **a** shows the relationship under least-direct-cost scenarios, and panel **b** shows the relationship under zero-emissions scenarios. Different types of generation are plotted in different colors.

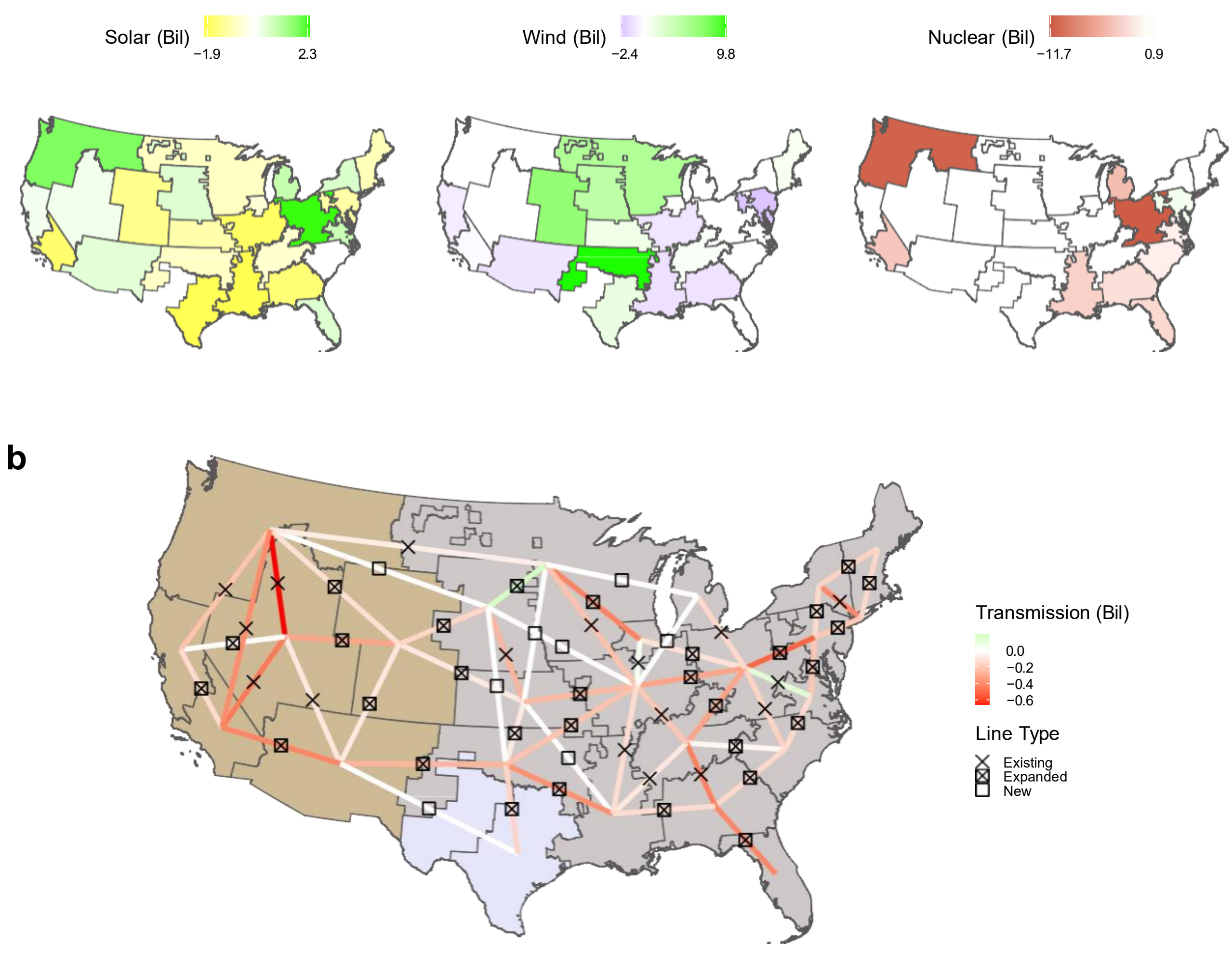

**Figure 6: Rent changes when going from existing to optimal transmission in zero-carbon scenarios.** Each graph shows the rent change to a resource when going from the ZE scenario to the ZO scenario. The rents accrue to infra-marginal wind and solar resources that are more valuable than marginal sources, existing nuclear facilities (costs are assumed sunk), and constrained transmission resources that receive surplus congestion rents. Panel **a** shows transmission expansion benefits solar and wind producers in regions unusually rich in these resources relative to their neighbors, while reducing rents in neighboring areas; it also shows that existing nuclear typically loses with transmission expansion since it becomes less competitive. Panel **b** shows the change in rents to transmission lines. Expanded transmission links typically lose rent as congestion charges decline. New transmission links have zero rent because expansion is optimized. Existing but non-expanded lines also tend to lose rent as congestion charges decline. The rare exceptions where transmission rents rise are cases where existing transmission is marginally overbuilt and losses are reduced.

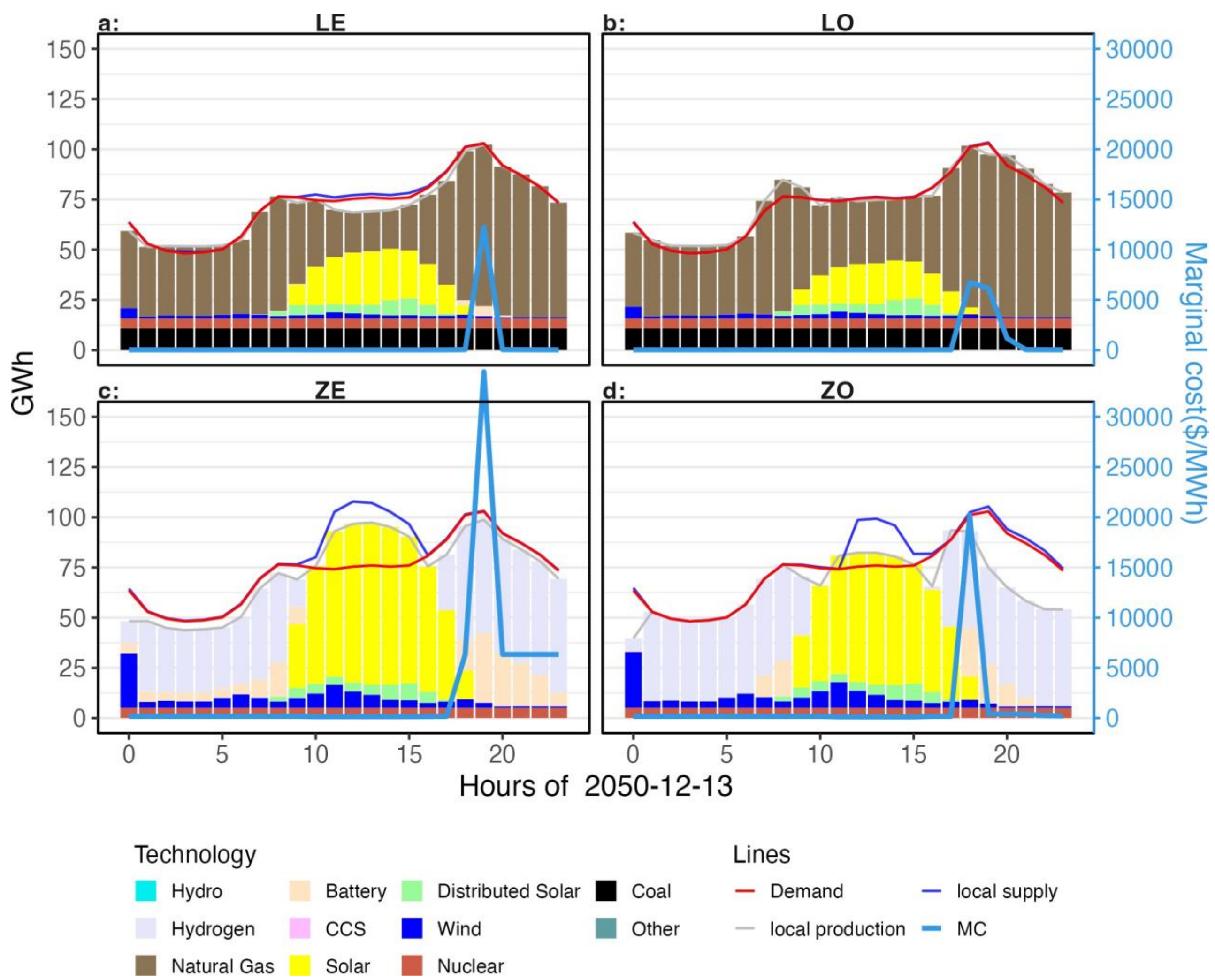

**Figure 7: Hourly generation, dispatch, transmission, and marginal cost in TRE during the most costly sample day.** The graphs show hourly dispatch, inflows, outflows, and marginal cost in four scenarios for the TRE region, the Eastern part of ERCOT in Texas, on the sample day with the highest demand-weighted average marginal cost. Panels **a** and **b** show the least-direct-cost systems under existing and optimized transmission (LE and LO), and panels **c** and **d** show the zero-emissions systems under existing and optimized transmission (ZE and ZO).

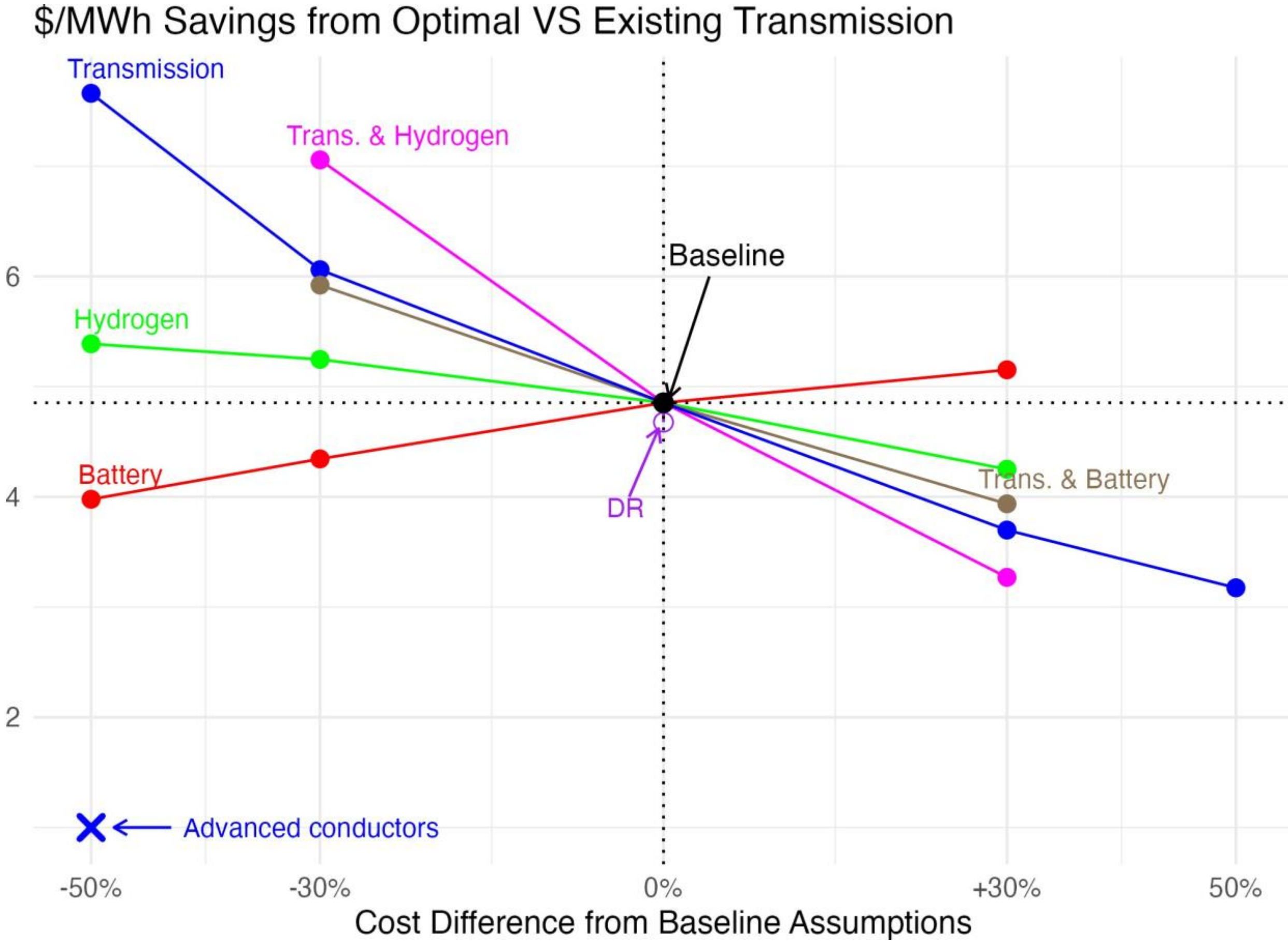

**Figure 8: The value of transmission expansion in a zero-emissions system with demand response and alternative cost assumptions.** The graph shows the value of transmission, measured as the difference in average load-weighted marginal cost ($/MWh) between optimized and existing transmission in a zero-emissions system (ZO−ZE). The baseline refers to the difference between ZE and ZO under NREL "Mid" cost assumptions reported in other figures and tables. The "DR" point holds all costs the same, but assumes 10% of demand in each hour can be shifted costlessly to another hour within the same day, but limits increases in any hour to 20% of baseline load. The other points and lines show how the value of optimal transmission expansion changes as battery costs, transmission costs, or hydrogen costs vary from the baseline. A complete reporting of costs under these alternative cost scenarios is reported in Table 2.

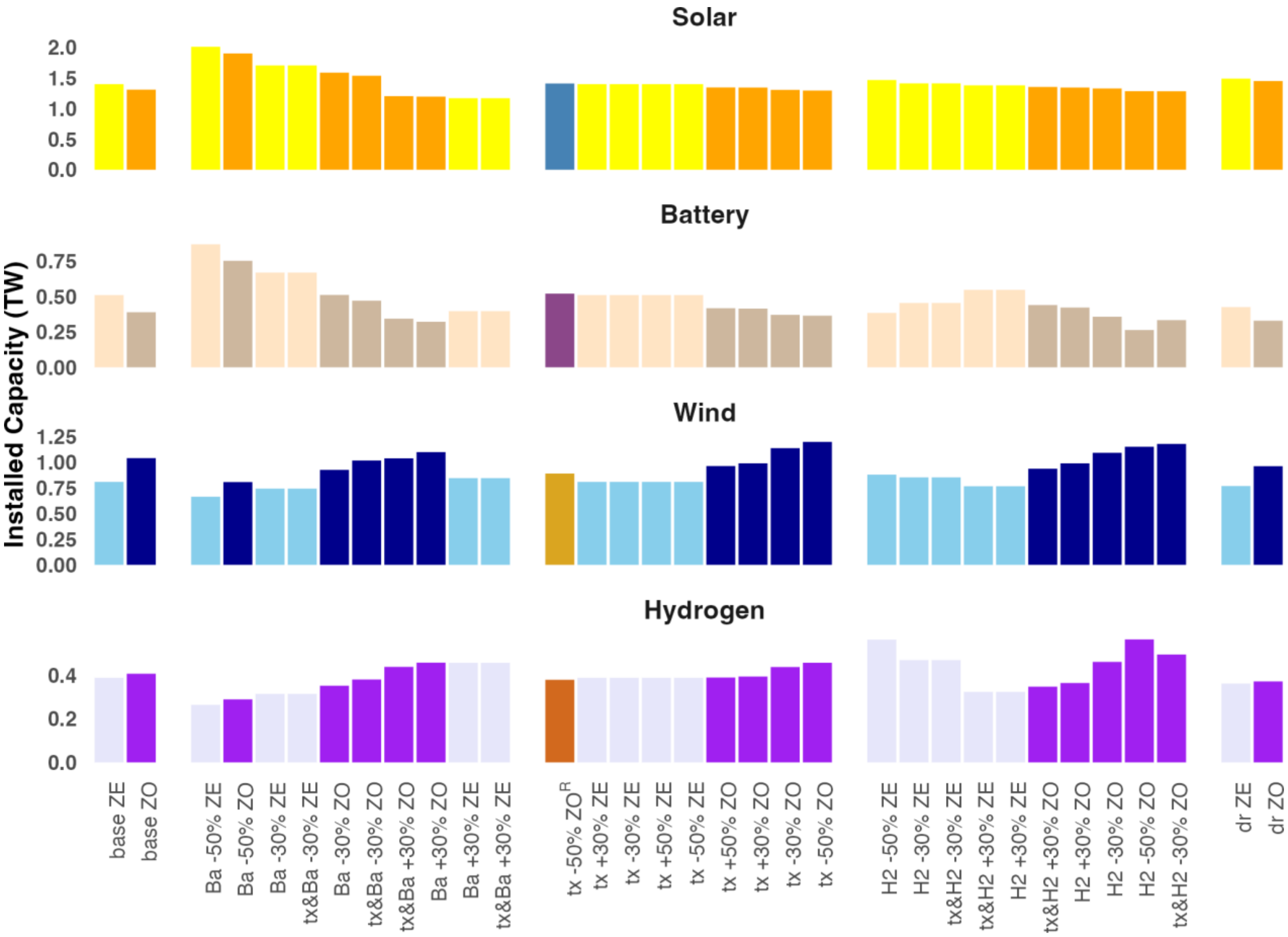

**Figure 9: Installed capacity by technology and alternative cost scenario.** The graph depicts variations in solar, battery, wind, and hydrogen capacities across alternative cost and transmission scenarios summarized in Table 2. All scenarios achieve zero emissions. The two scenarios on the far left correspond to the baseline ZE and ZO cases reported elsewhere and serve as reference points for scenarios with infrastructure costs above or below NREL "Mid" projections. The first group of ten bars to the right of the baseline cases represents changes in battery costs and combined battery+transmission costs. The second group of nine bars shows changes in transmission costs only, which affect the generation mix only in the ZO cases. Off-color bars within this group correspond to the "advanced conductors" scenario. The third group of ten bars reflects changes in combined hydrogen+transmission costs, and the final two bars show the effects of limited demand response (10% within-day shifting). Within each group, scenarios are ordered from the highest to the lowest solar capacity. Except for the advanced conductors case, darker-shaded bars represent scenarios with optimal transmission, while lighter-colored bars indicate existing transmission.

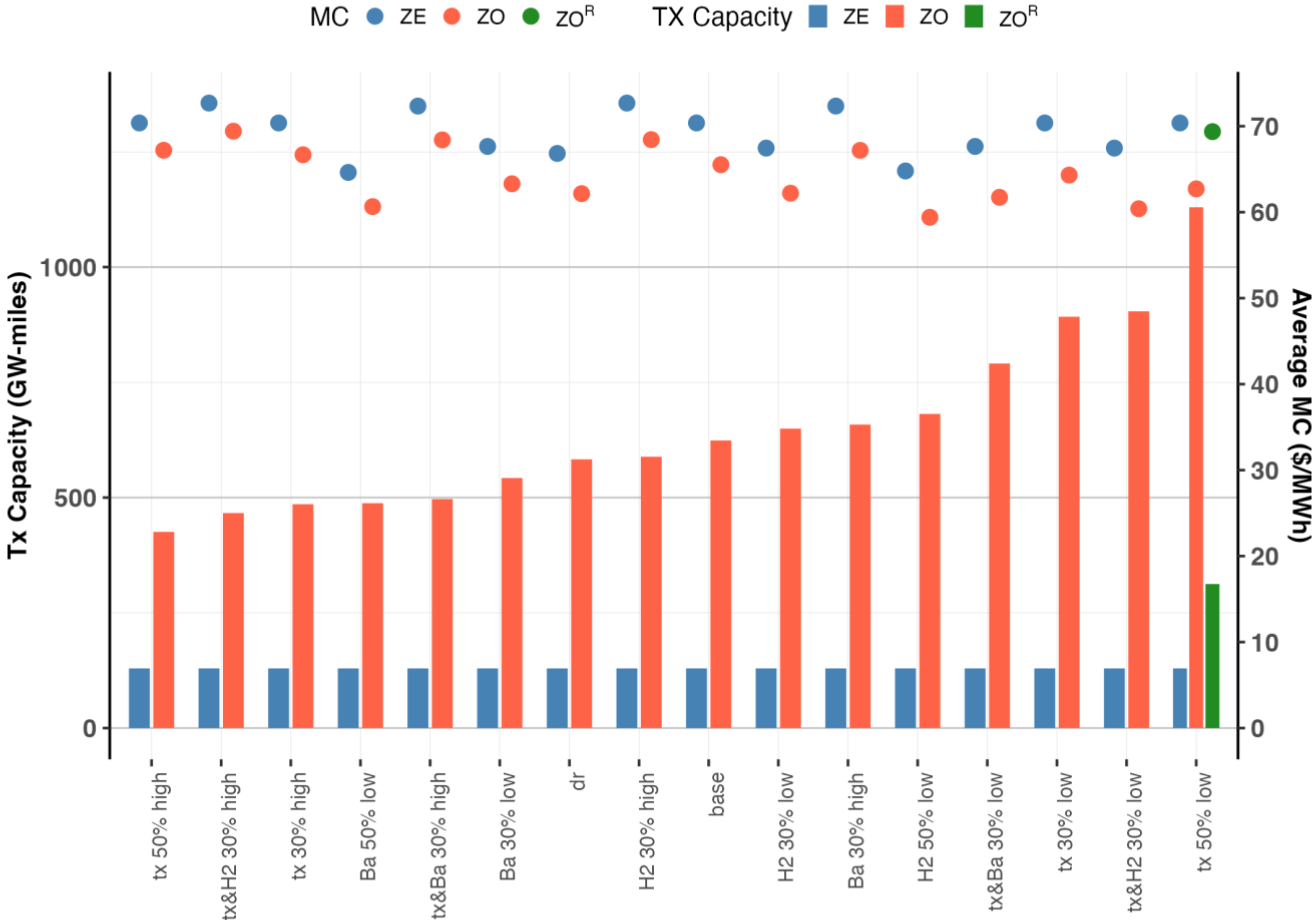

**Figure 10: Installed transmission capacity and load-weighted average marginal cost by alternative cost scenario.** The graph shows total interregional transmission capacity (GW) for zero-emissions scenarios under the alternative cost assumptions outline in section 3, each paired with load-weighted average marginal cost. The red bars are ZE scenarios with transmission constrained at existing capacities; the blue bars show unconstrained transmission optimization, and the single green bar shows the "advanced conductors" scenarios with half the baseline transmission cost per GW-mile, but with expansion constrained to a maximum of two times current transmission along each connection. Scenarios are ordered by the amount of transmission in the unconstrained ZO case.

# Optimal transmission expansion modestly reduces decarbonization costs of U.S. electricity

## Appendix

Here we describe the mathematical structure of Switch in more detail and present additional results, with an emphasis on comparing the least-direct-cost (LX) scenarios with the *socially optimal* (SX) scenarios that minimize costs subject to a carbon emissions price of $190 per ton rather than forcing zero emissions. This price leads to 90 to 95% emissions reductions relative to 2022 levels in most cases, even with a 75% growth in electricity production. We refer to these scenarios as socially optimal because they equate the marginal cost of abatement to the Environmental Protection Agency's marginal social cost of emissions (i.e., the estimated marginal benefit of abatement).

## Switch Formulations

Here we provide a brief overview of the mathematical formulation of Switch 2.0, the model used for our research. More complete documentation of the software can be found in Johnston et al. (2019). Switch 2.0 has a modular architecture that reflects the modularity of actual power systems. Most power system operators follow rules that maintain an adequate supply of power, and most individual devices are not concerned with the operation of other devices. Similarly, core modules in Switch define spatially and temporally resolved balancing constraints for energy and reserves, and an overall social cost. Separate modules represent components such as generators, batteries, or transmission links. These modules interact with the overall optimization model by adding terms to the shared energy and reserve balances and the overall cost expression. Switch 2.0 supports co-optimization of multiple investment periods, but we have omitted those definitions here, since we use a single stage for this study (2041-2050). We have likewise omitted details on spinning reserves and unit commitment, which were not used for this study.

### Objective function

The objective function is defined by the **financials** module. It minimizes the net present value of all investment and operation costs:

$$Min \left( \sum_{c^{\mathrm{f}} \in C^{\mathrm{fixed}}} c^{\mathrm{f}} + \sum_{t \in T_p} w_t^{\mathrm{year}} \sum_{c^{\mathrm{v}} \in C^{\mathrm{var}}} c_t^{\mathrm{v}} \right) \tag{1}$$

Function 1 sums over sets of fixed costs $C^{\mathrm{fixed}}$ and variable costs $C^{\mathrm{var}}$. Each fixed cost component $c^{\mathrm{f}} \in C^{\mathrm{fixed}}$ is a Pyomo object, indexed by investment period and specified in units of $/year. This object may be a variable, parameter, or expression (calculation based on other components). The term $c^{\mathrm{f}}$ is the fixed cost that occurs during our study period. Each variable cost component $c^{\mathrm{v}}$ is indexed by timepoint ($t$) and specified in units of $/hour. Modules add components to the fixed and variable cost sets to represent each cost that they introduce. Hence, the exact equation will depend on which modules are selected by the user. In most runs, fixed costs in $C^{\text{fixed}}$ include capital repayment for investments at a fixed financing rate over the lifetime of each asset, sunk costs from existing infrastructure, as well as fixed operating and maintenance (O&M) costs. Variable costs in $C^{\text{var}}$ typically include fuel costs and variable O&M. The weight factor $w_t^{\text{year}}$ scales costs from a sampled timepoint to an annualized value.

## Operational Constraints

*Power Balance:* Specifies that power injections and withdrawals must balance during each time point $t$ in each zone $z$. As with the objective function, plug-in modules add model objects to $P^{\text{inject}}$ and $P^{\text{withdraw}}$ to show the amount of power injected or withdrawn by each system component during each timepoint. For this study, production components include renewable and conventional generators, batteries, hydrogen fuel cells, and inbound transmission flows. Withdrawals include customer loads, battery charging, hydrogen electrolysis and refrigeration, and outbound transmission flows.

$$\sum_{p^{\text{i}} \in P^{\text{inject}}} p^{\text{i}}_{z,t} = \sum_{p^{\text{w}} \in P^{\text{withdraw}}} p^{\text{w}}_{z,t}, \qquad \forall z \in Z, \forall t \in T \tag{2}$$

*Dispatch:* Power generation from a source $g$ (e.g., a power plant) must fall below its installed capacity $K^{\text{G}}_g$ during time point $t$ multiplied by a capacity factor $\eta_{g,t}$, that may vary with exogenous factors like solar radiation or wind speed.

$$0 \leq P_{g,t} \leq \eta_{g,t} K^{\text{G}}_g \qquad \forall g \in G, \forall t \in T \tag{3}$$

*Transmission Flows:* Transmission flows $F_{\ell,t}$ along a corridor $\ell$ are constrained by the installed capacity $K^{\text{L}}_{\ell}$ on that corridor. Additional constraints (not shown) define the flow out of a corridor to take a smaller value than flow into the corridor, reflecting transmission losses.

$$0 \leq F_{\ell,t} \leq K^{\text{L}}_{\ell}, \qquad \forall \ell \in L, \forall t \in T_p \tag{4}$$

*Power System Construction:* Eqs. (5) and (6) define installed capacity for generation projects $K^{\text{G}}_g$ and transmission lines $K^{\text{L}}_{\ell}$ as the sum of capacity additions during the study ($B^{\text{G}}_g$ or $B^{\text{L}}_{\ell}$) and preexisting capacity of the same type in the same location ($k^{\text{G}}_g$ or $k^{\text{L}}_{\ell}$).

$$K^{\text{G}}_g = B^{\text{G}}_g + k^{\text{G}}_g, \qquad \forall g \in G \tag{5}$$

$$K^{\text{L}}_{\ell,p} = B^{\text{L}}_{\ell} + k^{\text{L}}_{\ell,y}, \qquad \forall \ell \in L \tag{6}$$

Some generation projects (the set $G^{\text{rc}}$) also have caps on installed capacity $\overline{k^{\text{G}}_g}$. These may be plants of a type that cannot be built in the future or renewable projects with limits on available land.

$$K^{G}_{g} \leq \bar{k}^{G}_{g}, \qquad \forall g \in G^{rc} \tag{7}$$

Additional terms define behavior of storage facilities (charging, state of charge, round-trip losses), fuel consumption (full load heat rate times power production) and hydrogen facilities (production and storage of hydrogen and conversion back to electricity). For complete details on Switch's mathematical formulation, see the Supplementary Material of Johnston et al. (2019) at https://ars.els-cdn.com/content/image/1-s2.0-S2352711018301547-mmc1.pdf.

# Supplemental Figures

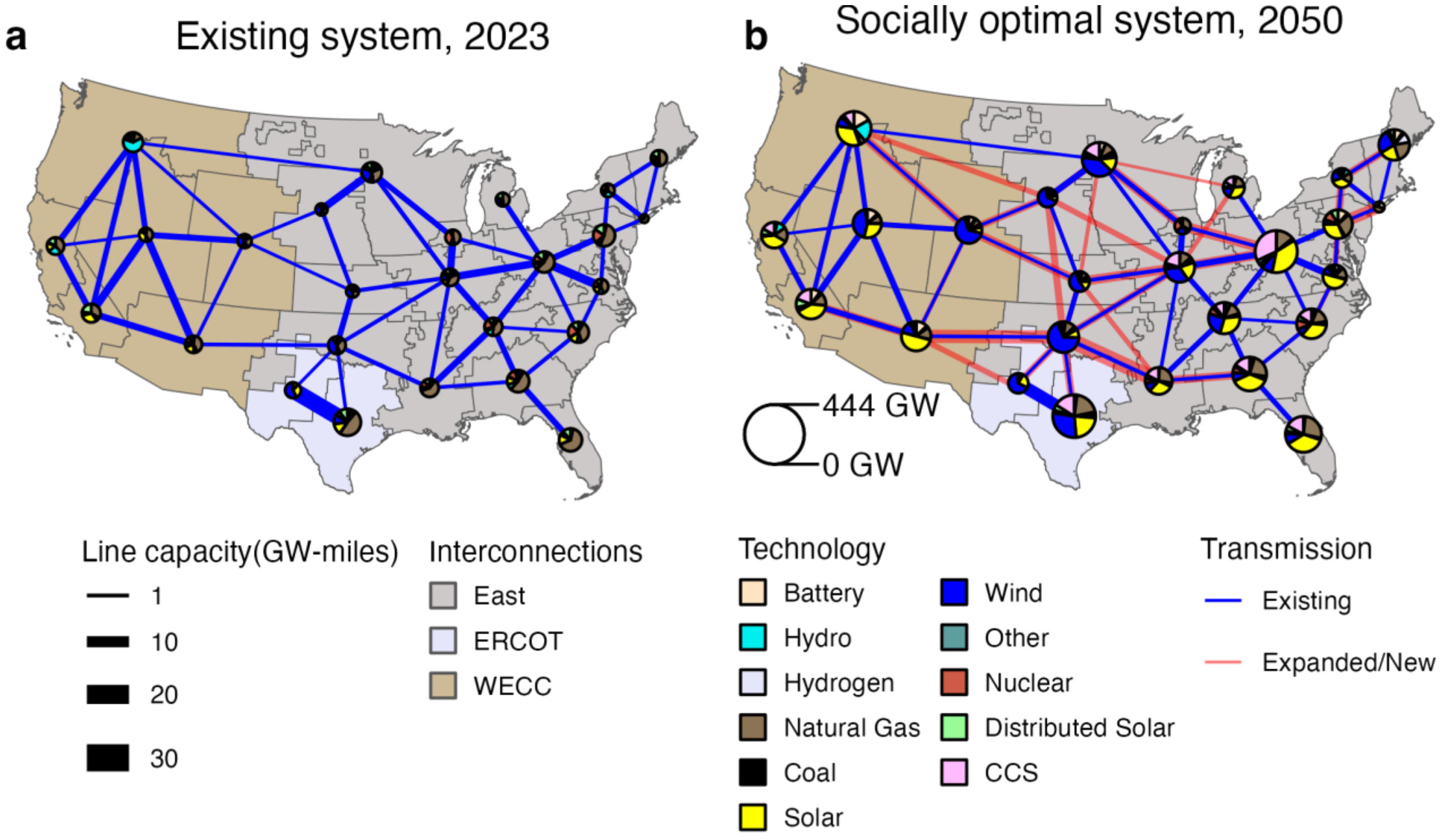

**Figure A1: Demand, generation, and transmission capacities in existing and idealized carbon-priced electricity systems.** Panel **a** shows 2023 generation capacities and inter-regional transmission. Panel **b** shows an optimized, carbon-priced system for 2050 without constraints on the generation mix or transmission expansion. It is important to note the substantial difference in scale of generation capacities for 2023 and 2050.

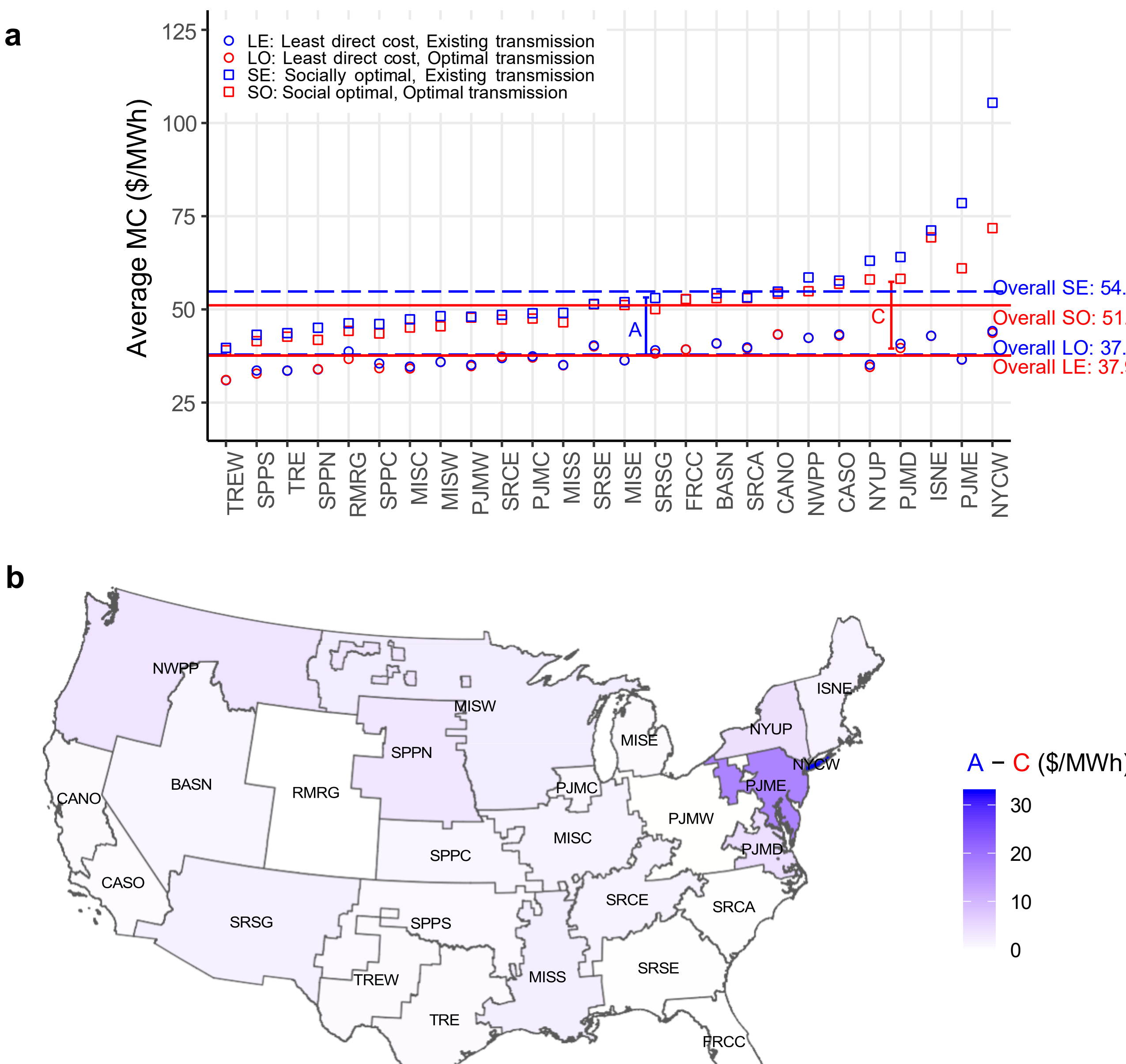

**Figure A2: Cost per MWh for different emissions and transmission scenarios.** Panel **a** shows the demand-weighted average marginal cost for each region in 2050 in four scenarios, least-direct-cost (circles) and carbon-priced scenarios (squares), each with existing (blue) and optimized (red) transmission. Comparing square to triangles of the same color gives the region's cost of decarbonization, with blue indicating the cost without transmission expansion (difference A) and red indicating the cost with optimized transmission expansion (difference C). Comparing the same shapes of different colors gives the net savings from expanded transmission. Panel **b** shows a map of the difference in differences (A-C): the cost savings ($/MWh) from optimizing transmission under full decarbonization relative to using only existing transmission.

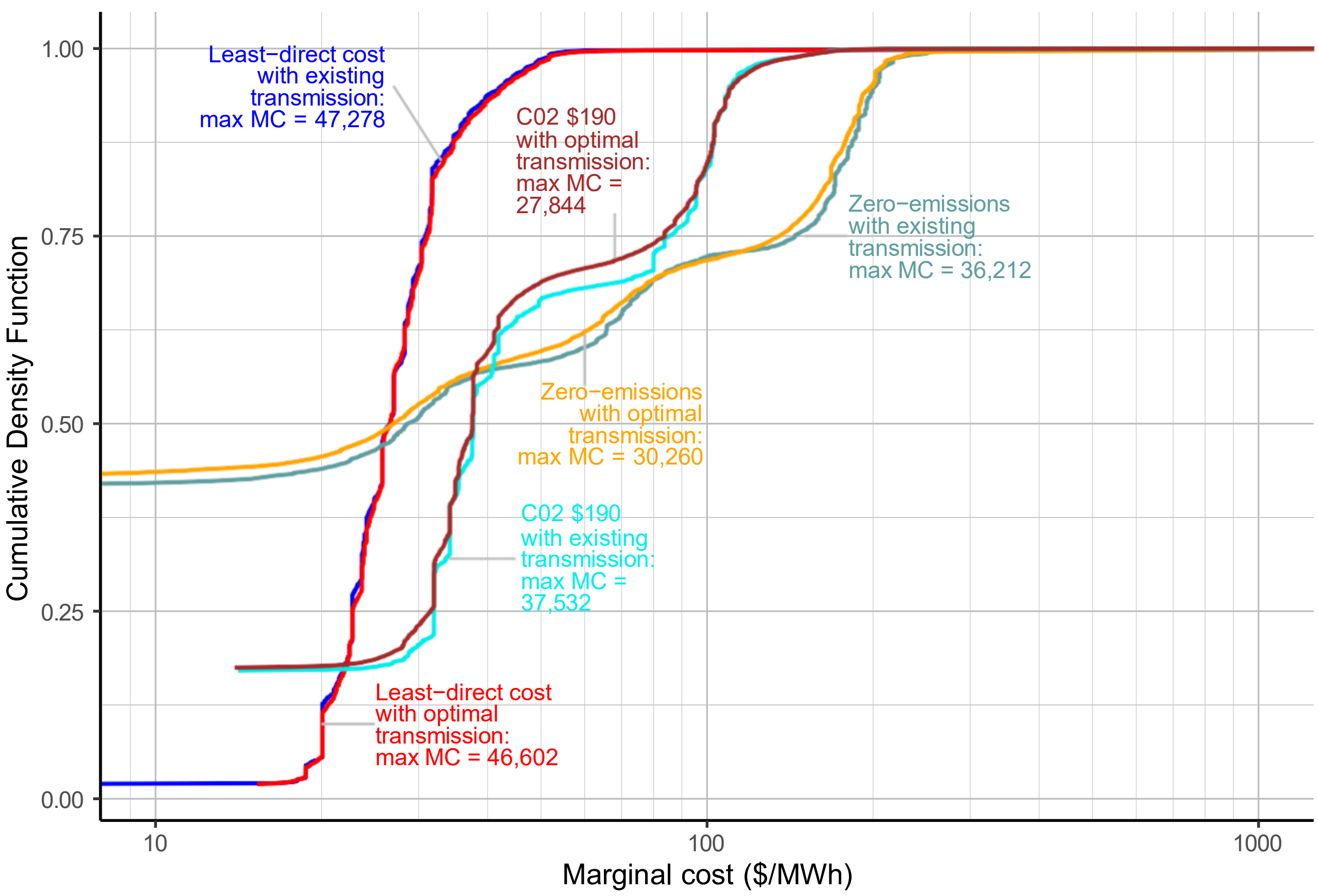

**Figure A3: Distribution function of marginal cost.** This graph shows the distribution functions or (cumulative density functions) for hourly marginal cost across all MWh in all continental U.S. regions in 2050. To construct these, we account for the MWh of demand in each region/hour. Six scenarios are depicted: least-direct-cost, with existing and optimal transmission (LE & LO), zero-emissions with existing and optimal transmission (ZE & ZO), and carbon-priced, with an assumed price of $CO_2$ emissions of $190 per ton, which achieves roughly 89% reduction of emissions from the electricity sector relative to 2022. All scenarios assume 74% demand growth relative to 2022 and conservative projections from NREL-ATB.

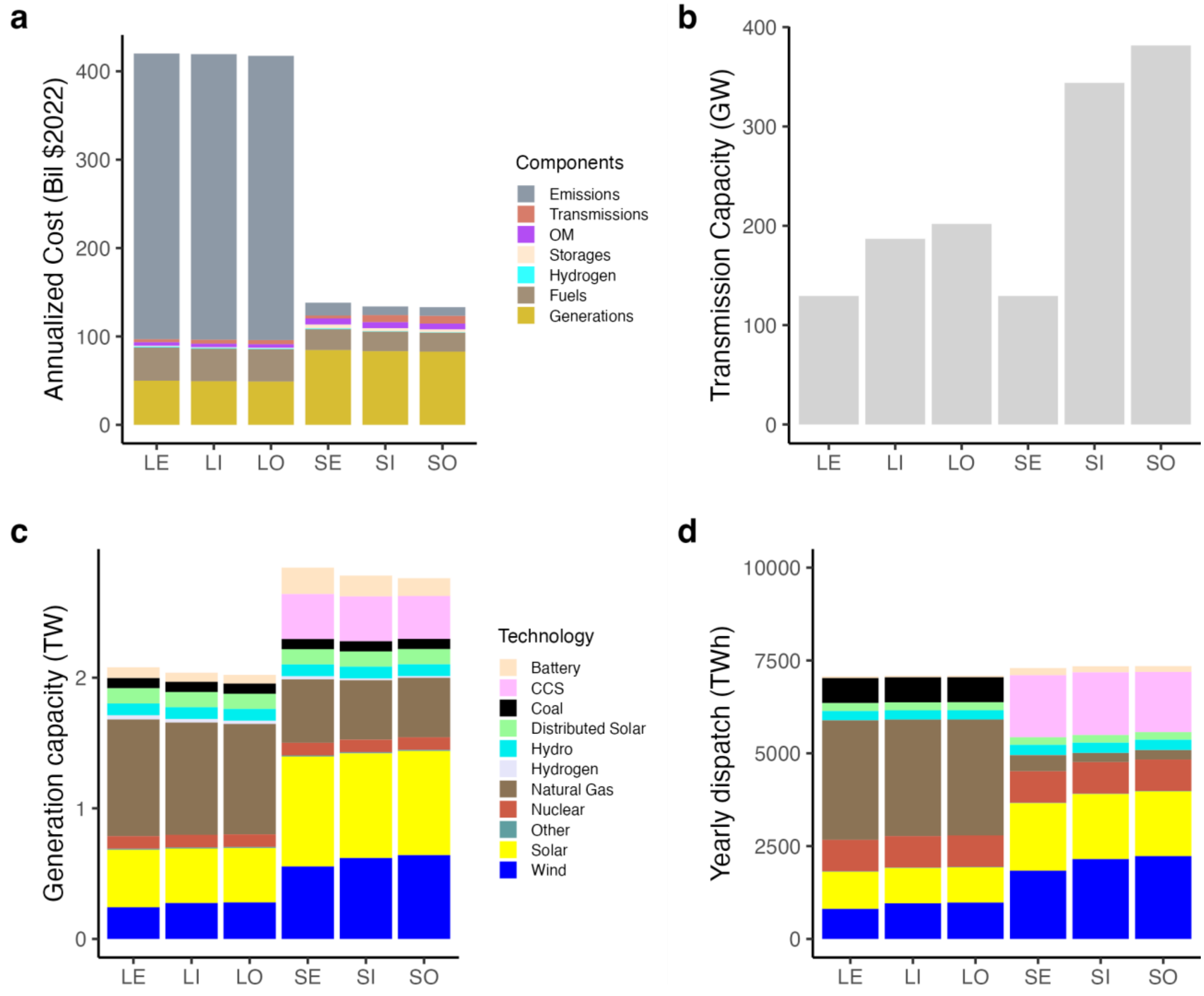

**Figure A4: Comparing component costs and capacities.** The graphs compare costs, generation mixes, and transmission capacities across six scenarios, LE (least-direct-cost with existing transmission), LI (least-direct-cost with optimal within-interconnect transmission), LO (least-direct-cost with optimal fully-optimized transmission), SE (carbon-priced with existing transmission), SI (carbon-priced with optimal within-interconnect transmission), and SO (carbon-priced with optimal fully-optimized transmission). Panel a shows broadly categorized cost components; panel b shows transmission capacity in each scenario (GW); panel c shows generation capacities (TW) in each scenario; and panel d shows the share of dispatch (source of energy consumed) in each scenario.

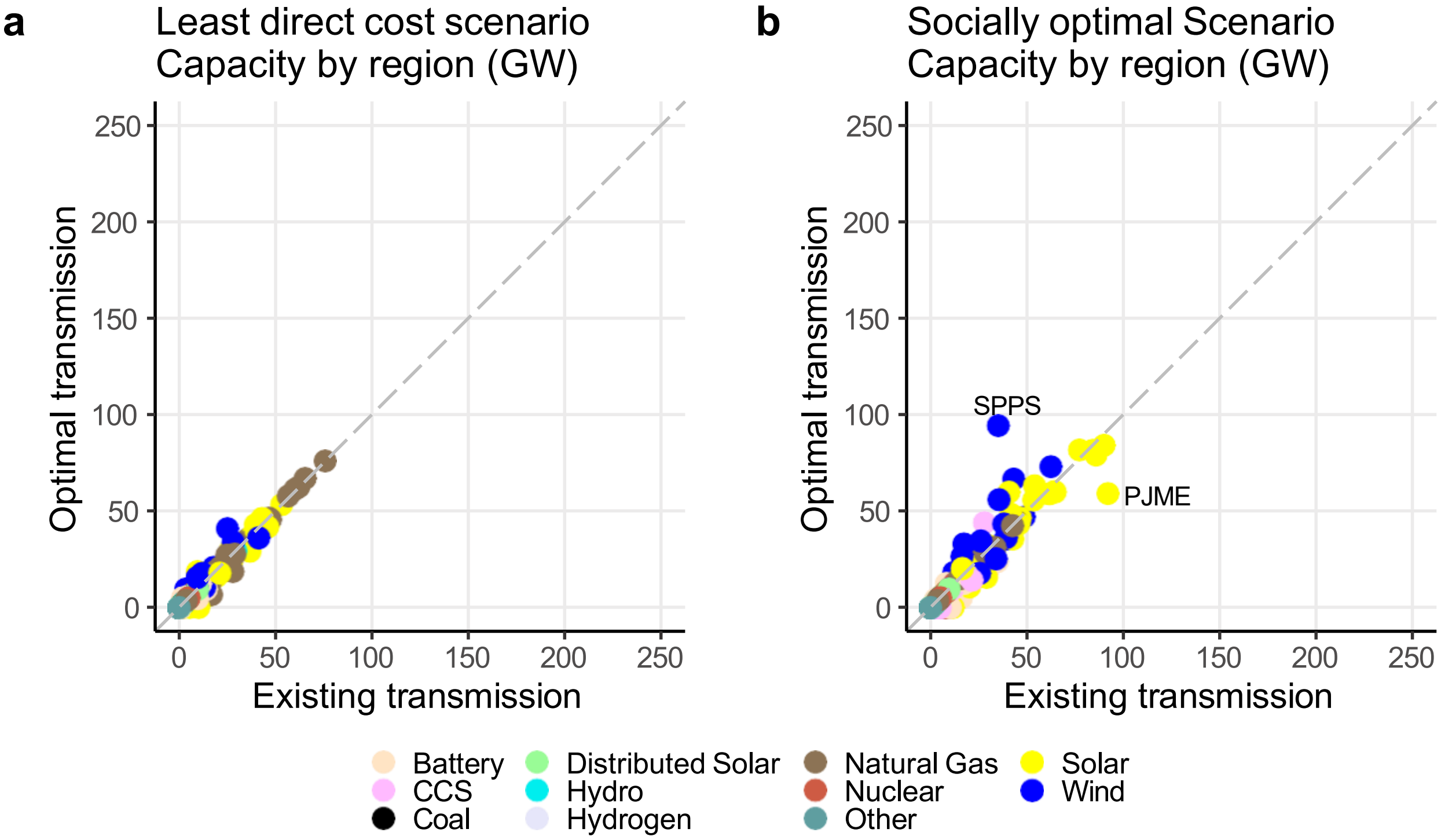

**Figure A5: Comparison of regional capacities across transmission scenarios.** These scatter plots show how transmission influences the mix of generation capacities across regions. Each panel shows optimized region-level generation capacities under existing transmission plotted against generation capacities under fully optimized transmission. Panel **a** shows the relationship under least-direct-cost scenarios, and panel **b** shows the relationship under carbon-priced scenarios. Different types of generation are plotted in different colors.

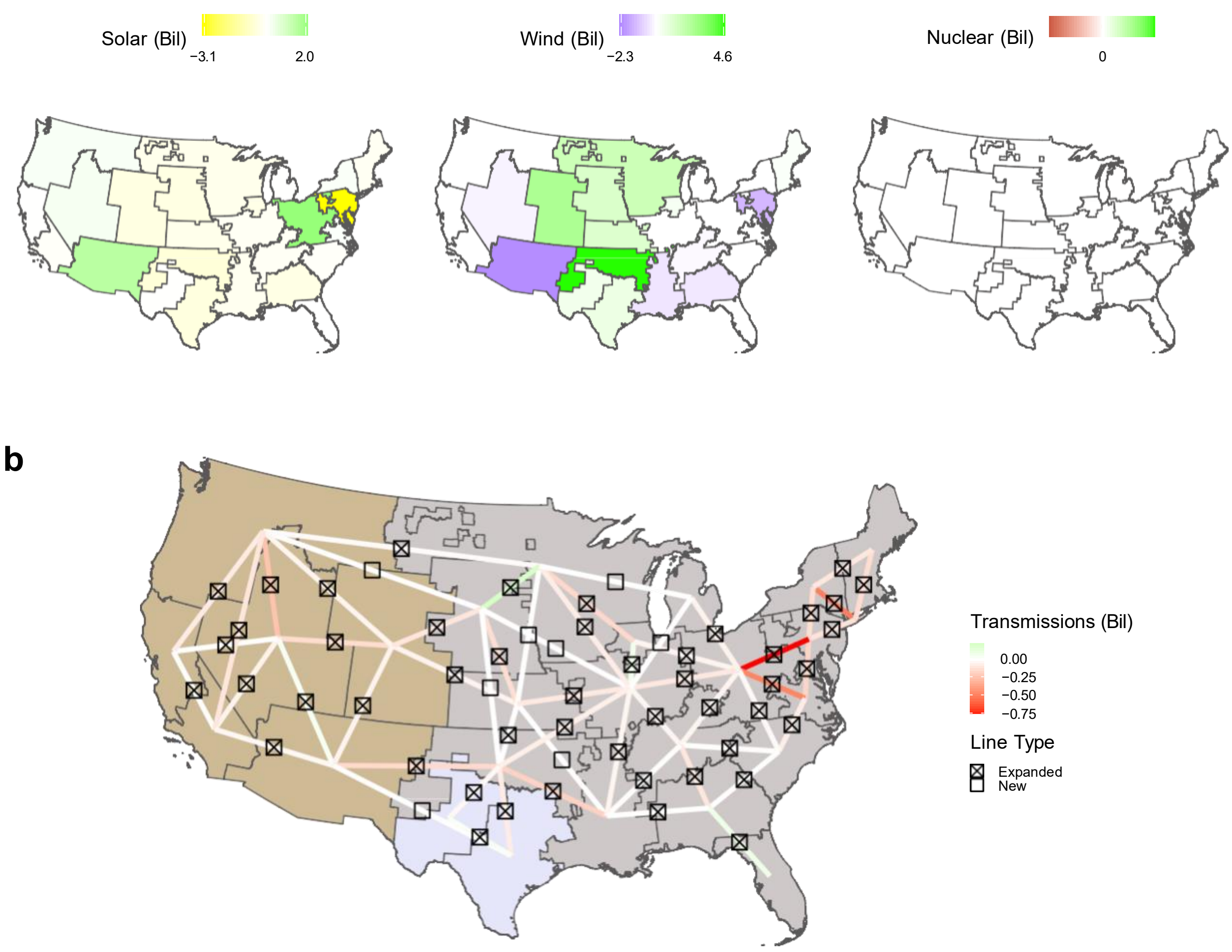

**Figure A6: Rent changes when going from existing to optimal transmission in carbon-priced scenarios.** Each graph shows the change in rents to a resource when going from the SE scenario to the SO scenario. The rents accrue to infra-marginal wind and solar resources that are more valuable than marginal sources, to existing nuclear facilities (costs are assumed sunk), and to constrained transmission resources that receive surplus congestion rents. Panel **a** shows transmission expansion benefits solar and wind producers in regions unusually rich in these resources while hurting producers in other regions; it also shows that existing nuclear generally gains with transmission expansion since it can enjoy higher capacity factors. Panel **b** shows the change in rents to transmission lines when going from the SE scenario to the SO scenario.

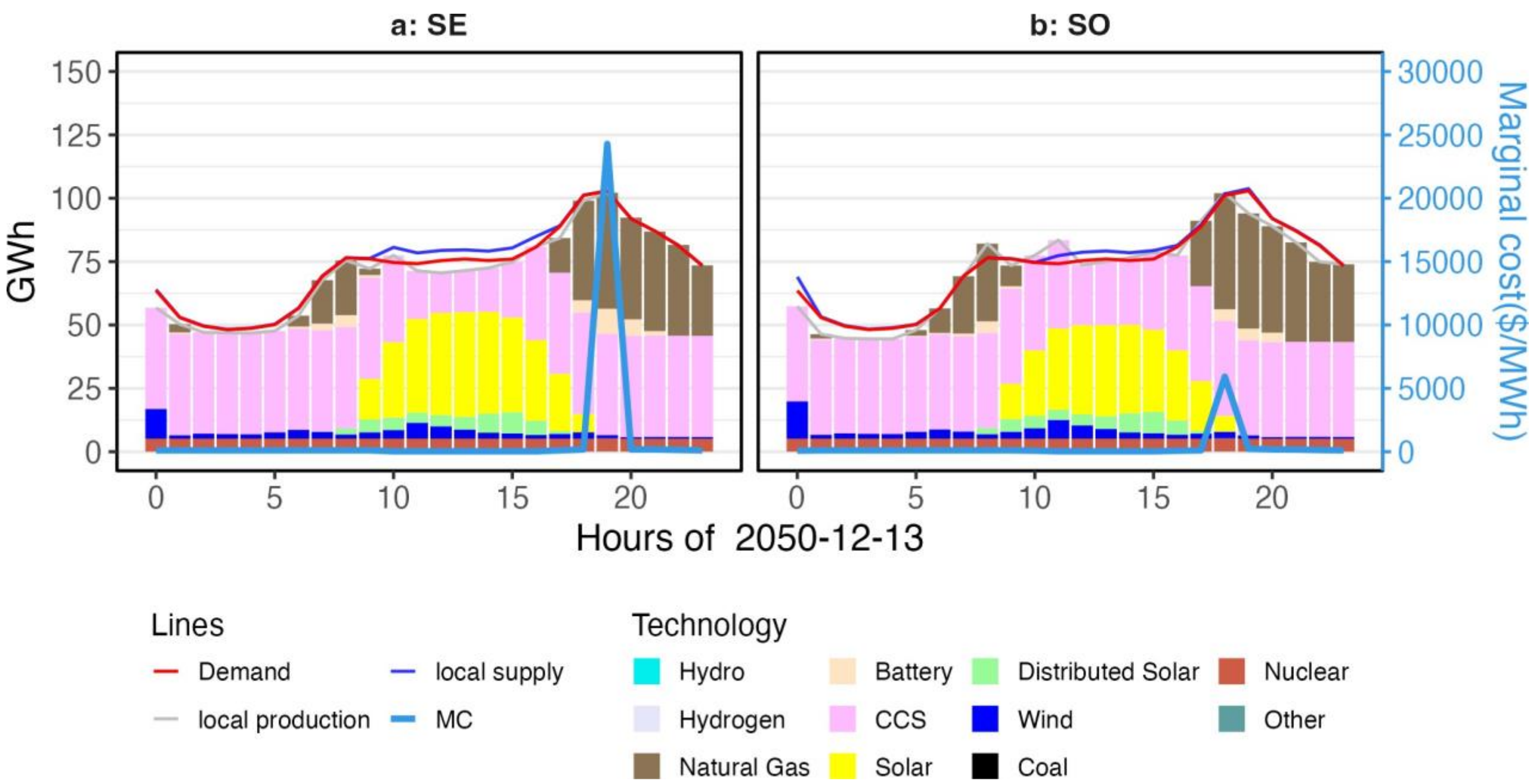

**Figure A7: Hourly generation, dispatch, transmission, and marginal cost in TRE during the most costly sample day.** The graphs show hourly dispatch, inflows, outflows, and marginal cost in four scenarios for the TRE region, which is the Eastern part of ERCOT in Texas, on the sample day with the highest demand-weighted average marginal cost – Dec 13th, 2050. Panels **a** and **b** show the carbon-priced systems under existing and optimized transmission (SE and SO).

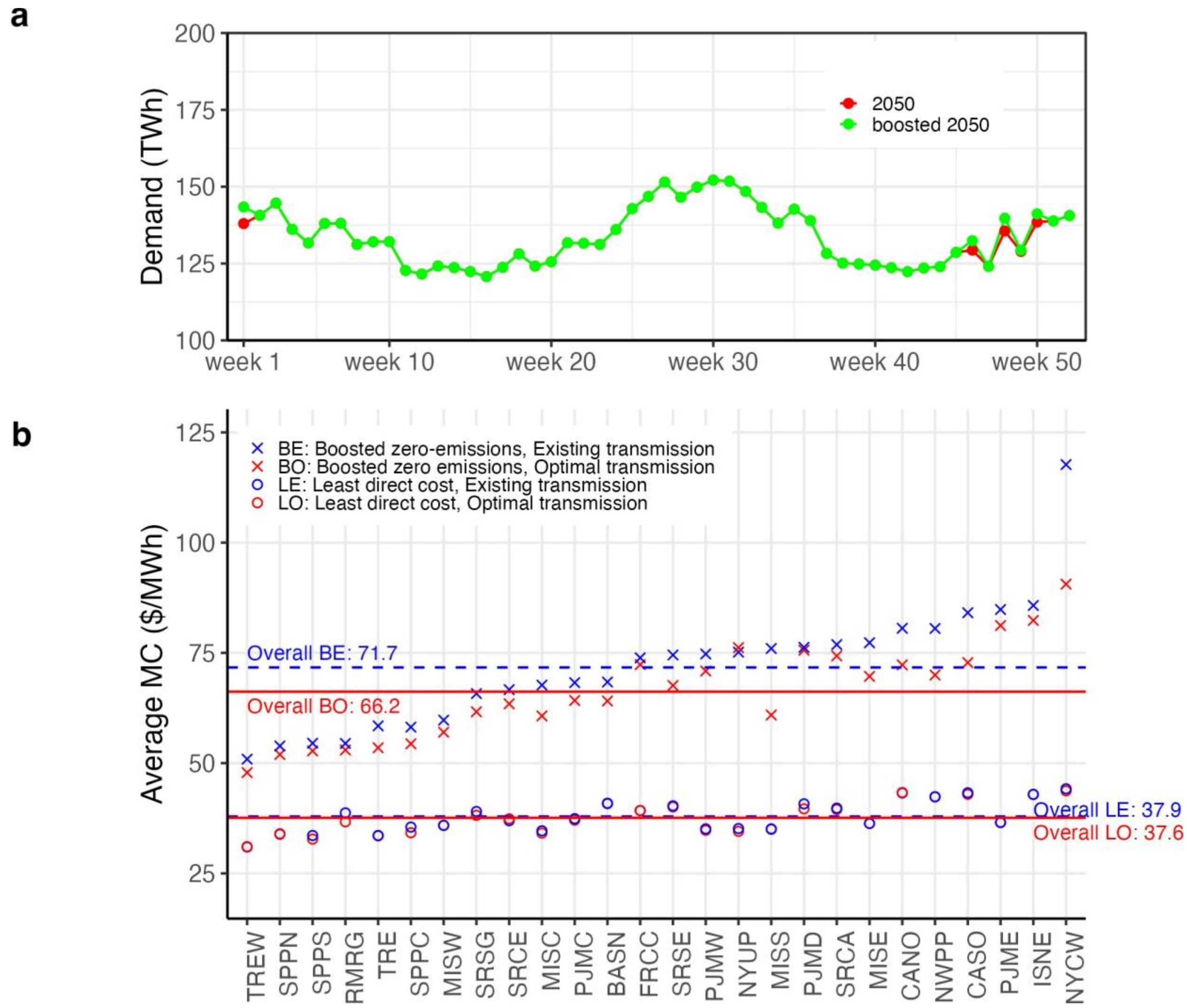

**Figure A8: Demand and marginal cost of original and boosted 2050 system.** Panel **a** shows the weekly demand of 2050 and the boosted system. The boosted system increases 25% of demand for the toughest 3 consecutive days for each regions. Panel **b** shows the demand-weighted average marginal cost for each region in 2050 in four scenarios, least-direct-cost (circles) and boosted zero-emission scenarios (crosses), each with existing (blue) and optimized (red) transmission.

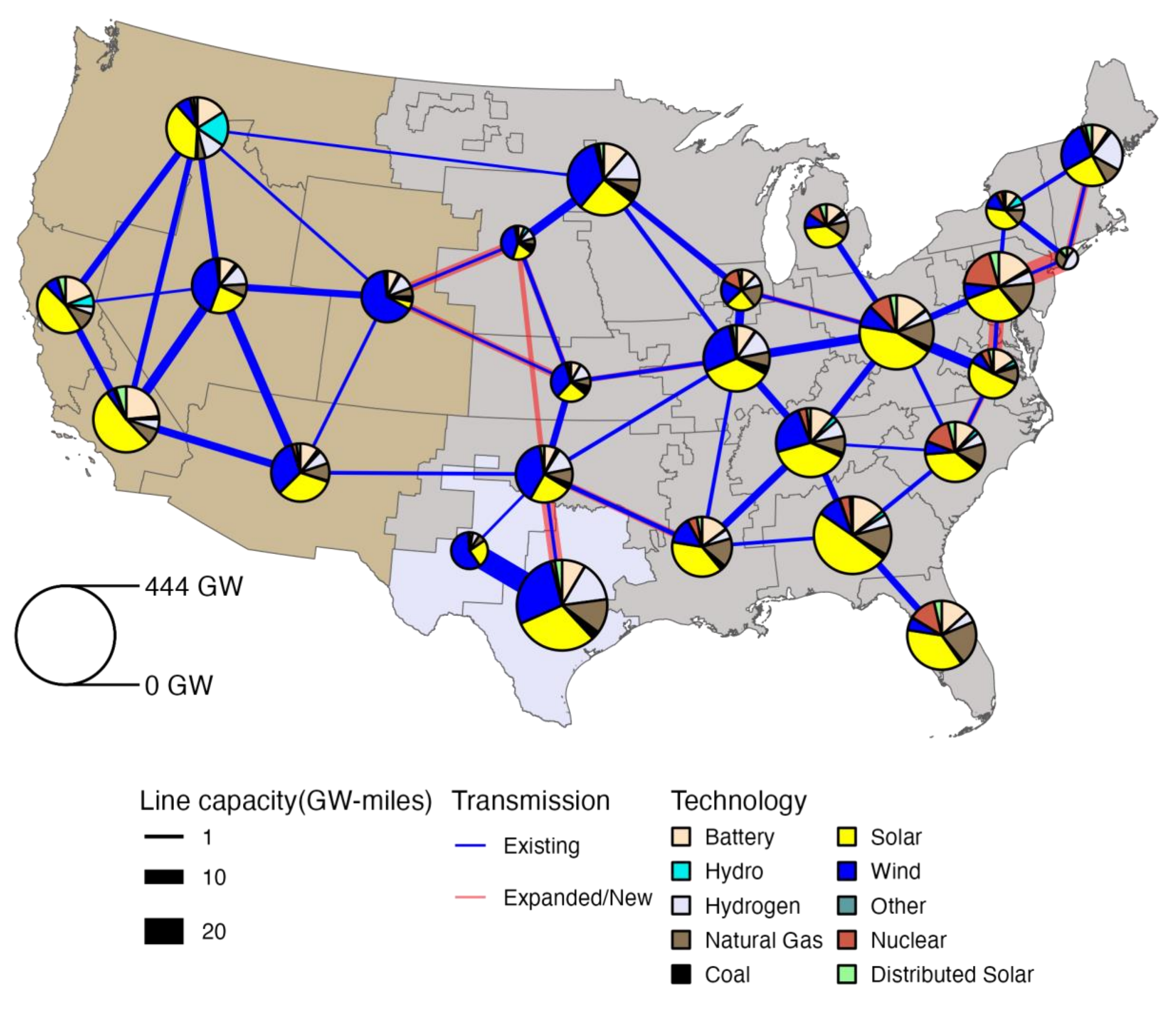

**Figure A9: Generation and transmission capacities in Zero-emissions, with transmission expansion limited to 25% of existing (ZE+).**